\documentclass[submission,copyright,creativecommons]{eptcs}
\usepackage{etex} 
\providecommand{\event}{HAS 2014} 
\usepackage{breakurl}             

\usepackage[]{graphicx}
\newif\ifignore 
\ignorefalse
\newcommand{\auxproof}[1]{
  \ifignore\mbox{}\newline
  \textbf{BEGIN: AUX-PROOF} \dotfill\newline
         {#1}\mbox{}\newline
         \textbf{END: AUX-PROOF}\dotfill\newline
         \fi}

\usepackage{times}

\usepackage{amsthm,algorithm,algorithmic}
\usepackage{fancybox,amssymb,amstext,amsmath,stmaryrd,wasysym,cite,proof,mathtools}
\usepackage[english]{babel}
\usepackage[all]{xy}
\usepackage{xspace}
\usepackage{epstopdf}
\allowdisplaybreaks[1] 

\xyoption{v2}
\xyoption{curve}
\xyoption{2cell}
\SelectTips{cm}{}  
\UseAllTwocells
\SilentMatrices
\def\labelstyle{\scriptstyle}

\newdir{ >}{{}*!/-8pt/@{>}}  
\newdir{|>}{%
  !/1.6pt/@{|}*:(1,-.2)@^{>}*:(1,+.2)@_{>}}
\newdir{pb}{:(1,-1)@^{|-}}
\def\pb#1{\save[]+<20 pt,0 pt>:a(#1)\ar@{pb{}}[]\restore}
\newcommand{\shifted}[3]{\save[]!<#1,#2>*{#3}\restore}
\usepackage{wrapfig}
\theoremstyle{plain}
\newtheorem{mytheorem}{Theorem}[section]
\newtheorem{mylemma}[mytheorem]{Lemma}
\newtheorem{myproposition}[mytheorem]{Proposition}

\theoremstyle{definition}
\newtheorem{mynotation}[mytheorem]{Notation}

\newtheorem{myexample}[mytheorem]{Example}
\newtheorem{mydefinition}[mytheorem]{Definition}

\newtheorem*{myproof}{Proof}
\newcommand{\myqed}{\hfill \ensuremath{\Box}}

\newcommand{\Sys}{{\mathcal{S}}}

\newcommand{\ImpCtrl}{{\mathbf{IMP_{Ctrl}}}}
\newcommand{\AssnCtrl}{{\mathbf{Assn_{Ctrl}}}}

\newcommand{\AExp}{{\mathbf{AExp}}}
\newcommand{\MExp}{{\mathbf{MExp}}}
\newcommand{\BExp}{{\mathbf{BExp}}}
\newcommand{\Cmd}{{\mathbf{Cmd}}}

\newcommand{\Fml}{{\mathbf{Fml}}}

\newcommand{\Skip}{{\mathtt{skip}}}
\newcommand{\If}[3]{{\mathtt{if}\;#1\;\mathtt{then}\;#2\;\mathtt{else}\;#3}}
\newcommand{\True}{{\mathtt{true}}}
\newcommand{\False}{{\mathtt{false}}}
\newcommand{\State}{{\Sigma}}

\newcommand{\N}{{\mathbb{N}}}

\newcommand{\R}{{\mathbb{R}}}
\newcommand{\B}{{\mathbb{B}}}

\newcommand{\Denote}[1]{{ \llbracket #1 \rrbracket}}
\newcommand{\Assn}[2]{{#1 \models #2}}
\newcommand{\NotAssn}[2]{{#1 \nvDash #2}}

\newcommand{\WPsyntax}[2]{{\mathsf{w} \llbracket #1,#2 \rrbracket }}

\newcommand{\SPsyntax}[2]{{\mathsf{s} \llbracket #1,#2 \rrbracket }}
\newcommand{\Defiff}{\stackrel{\text{def.}}{\Longleftrightarrow}}
\newcommand{\Execplant}{{\mathsf{execPlant}}}
\newcommand{\Sense}{{\mathsf{sense}}}
\newcommand{\Think}{{\mathsf{think}}}
\newcommand{\Act}{{\mathsf{act}}}

\newcommand{\CtrlCond}{{\pi_{\mathsf{C}}}}
\newcommand{\PlantCond}{{\pi_{\mathsf{P}}}}

\newcommand{\FA}{{\mathsf{FA}}}
\newcommand{\OneFA}{{1\text{-}\mathsf{FA}}}
\newcommand{\PreOneFA}{1\text{-}\mathsf{FA}^\mathsf{pre}}
\newcommand{\kFA}{{k\text{-}\mathsf{FA}}}
\newcommand{\init}{\mathsf{init}}
\newcommand{\final}{\mathsf{final}}
\newcommand{\PreBS}{1\text{-}\mathsf{BS}^{\mathsf{pre}}}
\newcommand{\BS}{{\mathsf{BS}}}
\newcommand{\cnt}{\mathsf{cnt}}

\newcommand{\Var}{\mathbf{Var}}
\newcommand{\Varthink}{\mathbf{Var}_{\mathsf{t}}}
\newcommand{\Varsense}{\mathbf{Var}_{\mathsf{s}}}
\newcommand{\Varact}{\mathbf{Var}_{\mathsf{a}}}
\newcommand{\ttrue}{\mathrm{t{\kern-0.8pt}t}}
\newcommand{\ffalse}{\mathrm{f{\kern-1.0pt}f}}
\newcommand{\Modes}{\mathsf{Modes}}

\newcommand{\VsenseSet}{\Varsense}
\newcommand{\VthinkSet}{\Varthink}
\newcommand{\Vthink}{{x_{\mathsf{t}}}}
\newcommand{\Vsense}{{x_{\mathsf{s}}}}
\newcommand{\Vact}{{x_{\mathsf{a}}}}
\newcommand{\rev}{\mathsf{rev}}

\newcommand{\aop}{\mathbin{\mathtt{aop}}}

\newcommand{\rop}{\mathrel{\mathtt{rop}}}

\newcommand{\place}{\underline{\phantom{n}}\,}

\begin{document}


\pagestyle{headings}  


\title{
  Input Synthesis for Sampled Data Systems by Program Logic
}

\author{
  Takumi Akazaki \qquad \qquad Ichiro Hasuo
  \institute{Department of Computer Science}
  \institute{The University of Tokyo, Japan}
  \email{ultraredrays@is.s.u-tokyo.ac.jp \quad ichiro@is.s.u-tokyo.ac.jp}
  \and
  Kohei Suenaga
  \institute{Graduate School of Informatics}
  \institute{Kyoto University, Japan}
  \email{ksuenaga@fos.kuis.kyoto-u.ac.jp}
}
\def\titlerunning{Input Synthesis for Sampled Data Systems by Program Logic}
\def\authorrunning{T. Akazaki, I. Hasuo \& K. Suenaga}





\maketitle

\begin{abstract}
  Inspired by a concrete industry problem we consider the \emph{input
  synthesis} problem for hybrid systems: given a hybrid system that is
  subject to \emph{input} from outside (also called \emph{disturbance}
  or \emph{noise}), find an input sequence that steers the system to the
  desired postcondition.  In this paper we focus on 
 \emph{sampled data systems}---systems in which a digital
 controller interrupts a physical plant in a periodic manner, a class commonly
 known in control theory---and furthermore assume that
a controller is given in the form of an  imperative program.
We develop a structural approach to input
 synthesis that
 features forward and backward reasoning in
 \emph{program logic} for the purpose of reducing a search space.
Although the examples we cover are limited both in size and in structure, 
experiments with a prototype
 implementation suggest potential of our program logic based approach.





\end{abstract}
%



%

%

\section{Introduction}
\label{section:introduction}



\emph{Cyber-physical systems (CPS)}---integration of digital control with
physical environments---are gaining yet more and more importance, with
cars, airplanes and all others controlled by computers. 
\emph{Hybrid systems} capture one of the crucial aspects of CPS,
by focusing on the combination of continuous \emph{flow} dynamics
and discrete \emph{jump} dynamics.
Quality assurance of hybrid systems is therefore a big concern in 
industry as well as in academia.


In this paper we study the \emph{input synthesis} problem of hybrid systems:
given a hybrid system that is subject to \emph{input} from
outside (also commonly called \emph{disturbance} or \emph{noise}), we
aim to find
an input sequence that steers the system to the desired postcondition. 
Our interest in input synthesis stems from the following concrete
problem; it was
provided by our research partner in car manufacturing industry as
a prototype of the problems they often encounter in their design process.

% ...
\begin{myexample}\label{example:TYTexample}
  In the system below in Fig.~\ref{fig:TYTexample}, 
  the controller interrupts the plant (a car) once every second
  and manages the velocity $v$ of the car.
  The controller chooses one mode $\tt m_{i}$ and the plant
  operates in that mode for one second, after which 
  the value of $v$ is fed back to the controller via the sensor.
  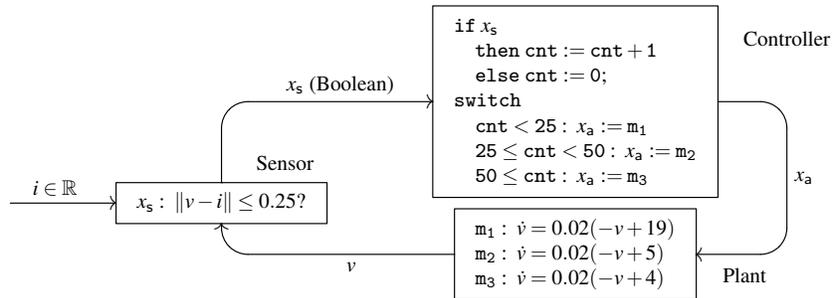
\begin{figure}
    \centering
    \begin{math}
      \scriptsize
      \def\labelstyle{\textstyle}
      \vcenter{\xymatrix@R=-.7em@C+2em{
          & & *+[F]{\begin{array}{l} 
              \tt if\; \mathit{x}_{\mathsf{s}}\\
              \quad \tt then\; cnt := cnt + 1\\
              \quad \tt else\; cnt := 0;\\
              \tt switch\; \\
              \quad \tt  cnt < 25:\; \mathit{x}_\mathsf{a}:= m_1\\
              \quad \tt  25 \leq cnt < 50:\; \mathit{x}_\mathsf{a}:= m_2\\
              \quad \tt  50 \leq cnt:\; \mathit{x}_\mathsf{a}:= m_3
          \end{array}}
          \shifted{10em}{3em}{\text{Controller}}
          \ar`r[dd]+/r10em/`[dd]^{x_\mathsf{a}}[dd]
          \\
            {}
            \ar[r]^*[r]{i\in \R \quad \quad}
            &
            *+[F]{
              \begin{array}{l} 
                \Vsense:\; \|v-i\|\le 0.25 ?
              \end{array}
            }
            \shifted{3em}{1.9em}{\text{Sensor}}   
            \ar`u[ru] [ru]^-{\Vsense \text{ (Boolean)}}
            \\
            & & *+[F]{\begin{array}{l} 
                {\tt m_1}:\; \dot{v}=0.02(-v + 19)\\
                {\tt m_2}:\; \dot{v}=0.02(-v + 5)\\
                {\tt m_3}:\; \dot{v}=0.02(-v + 4)
            \end{array}}
            \shifted{8em}{-1em}{\text{Plant}}
            \ar`l[lu]^-{v}[lu]
      }}
    \end{math}
    \abovecaptionskip = 0pt
    \caption{A hybrid system}
    \label{fig:TYTexample}
  \end{figure}%
  %
  The problem is to  come up with an initial state of the whole system
  together with an
  input sequence $i_0 \cdots i_{999}$, such that:
  \vspace*{-.8em}
  \begin{itemize}
  \item \textbf{(precondition)} the initial state satisfies $\cnt = 0$ and $x \in [-0.1, 0.1]$; and
  \item \textbf{(postcondition)} after 1000 seconds, the system satisfies $\cnt = 100$.
  \end{itemize}
\end{myexample}

The input synthesis problem can arise in many different contexts in
quality assurance of hybrid systems. One example is \emph{testing}: the
desired postcondition is the trigger for some countermeasure (e.g.\ a
fuse) against certain extremity (the countermeasure is outside the
model); and we seek for input (i.e.\ a test case) that drives the
system to activating the countermeasure. The input sequence thus
discovered in the model can be fed to the physical realization of the
system to see if the countermeasure works properly.

This paper contributes an algorithm for solving the input synthesis
problem. Its novelty is the use of \emph{program logic}: we make the
most of the structures expressed in the digital controller given in the
form of a program. In fact, a likely human effort for the  problem in Example~\ref{example:TYTexample}
is:
\begin{quote}
  $(*)$ ``for the system to have $\cnt = 100$ at
 time $k=1000$, the Boolean value $x_{\mathsf{s}}$ must be true from
 $k=900$ through $k=999$, and $\dotsc$'';
\end{quote}
this 
is nothing but reasoning in
program logic and is included in our proposed algorithm.

More specifically, we restrict our attention to a class of hybrid systems commonly called
\emph{sampled data systems}. One such system consists of a physical
\emph{plant}, a digital \emph{controller} that 
periodically interrupts the plant 
(for simplicity we assume a fixed interval), and a \emph{sensor} that
feeds  the state of the plant back to the controller.  This structural
assumption---restrictive yet realistic---allows us to
think of the
behaviors of such systems quite much as the semantics of programs, and 
enables forward and backward reasoning in program logic. 
In our algorithm for solving
the input synthesis problem, reasoning in program logic
(like the above $(*)$) contributes to the reduction of the search space.
Indeed our  prototype
implementation successfully solves the problem in
Example~\ref{example:TYTexample}.

\paragraph{Related Work}
The closest to the current work is one by Zutshi, Sankaranarayanan and
Tiwari~\cite{DBLP:conf/cav/ZutshiST12}, where they verify safety
properties 
of sampled data systems. Their model is more expressive, in that a plant
can autonomously change its modes without interruption by a controller.
While their goal is reachability analysis and is different from the
current paper's, their relational abstraction technique can be useful 
in our algorithm, too, in particular for the forward approximation
phase.

SMT-solver based
approaches~\cite{DBLP:conf/cade/GaoKC13,DBLP:conf/sefm/EggersRNF11} to
hybrid system analysis are related, too, especially in their emphases
on discrete jump dynamics rather than continuous flow. 
Their effectivity  in the input synthesis problem is not yet clear, though: the
only available implementation (that of
$\mathsf{dReal}$~\cite{DBLP:conf/cade/GaoKC13}) returned `unsat' to
Example~\ref{example:TYTexample}. 

More generally, an important feature of our modeling is that a digital
controller is given in the form of a program, unlike an automaton used in a
majority of existing work (including~\cite{DBLP:conf/cav/ZutshiST12,DBLP:conf/cade/GaoKC13}).
The contrast is comparable to the difference between the \emph{theorem
proving} (or \emph{type-based}) approach and \emph{software model
checking} in
program verification. While there have been 
results~\cite{KobayashiO09,DBLP:journals/toplas/NaikP08} that suggest these two
approaches are equivalent on a fundamental level, differences do
remain especially in applications. In our proposed algorithm it is an
advantage that we can exploit  rich structural information that is explicit
in a program in inferring impossibility ($\False$)  more quickly.

The backward search phase of our algorithm resembles a
\emph{membership question} addressed in the seminal
work by Alur et al.~\cite{DBLP:conf/rtss/AlurKV98}. Since our plant (flow) dynamics
is not necessarily linear, it is not easy to see how the results
in~\cite{DBLP:conf/rtss/AlurKV98} can be used in our
problem. 
They could nevertheless be applied to  meta-properties of the problem
such as complexity.




Fainekos and his colleagues have developed several techniques for
analyzing \emph{robustness} of hybrid systems.  Among them
is a tool called
S-Taliro~\cite{DBLP:conf/tacas/AnnpureddyLFS11}: it searches for 
a trajectory by optimization 
that relies on the continuous nature of the system dynamics. 
It is possible to encode the input synthesis problem into an input to
S-Taliro. However
 our  leading example
(Example~\ref{example:TYTexample}), of a jump-heavy nature, seems to fall out of the tool's 
focus (it timed out with a smaller problem of 15, not 1000, time units).

Several techniques for testing hybrid systems have been
proposed~\cite{DBLP:conf/soqua/BadbanFPT06,DBLP:journals/fmsd/DangN09,DBLP:conf/hybrid/JuliusFALP07,DBLP:conf/emsoft/AlurKRS08,EspositoTestCaseGeneration}.
Although they synthesize test cases and therefore seem similar to what
we do here, 
their goal is to meet
certain coverage criteria (such as star discrepancy
in~\cite{DBLP:journals/fmsd/DangN09}) and not to come up with input
that steers the system to a specific desired postcondition.

\auxproof{
  Input synthesis also serves for software testing, although most of the
  techniques proposed so far aim at meeting certain coverage criteria.
  The technique proposed by Charreteur and
  Gotlieb~\cite{DBLP:conf/issre/CharreteurG10} seems close to ours;
  their technique synthesizes inputs for programs written in Java
  bytecode.  Their technique is based on depth-first backward reasoning.
  They support heap structure that is essential for Java programs.
  However, because their target is software, they do not support
  continuous flow.
}

The current work is on  logical analysis of hybrid systems; and
in that respect it is close to Platzer's recent series of work
(see e.g.~\cite{DBLP:conf/lics/Platzer12a}) where dynamic logic is extended in a systematic way
so that it  encompasses continuous dynamics too. Also related is
the work~\cite{SuenagaH11ICALP,SuenagaSH13POPL} by some of the authors
where: flow is turned into jump with the help of \emph{nonstandard analysis}; 
and (discrete) program logic is applied as it is to hybrid systems.

\paragraph{Future Work}
In this paper we applied program logic to the specific problem of input
synthesis. We believe the technique have a greater potential and plan to 
look at other applications.

The current implementation can only handle continuous plants  of
dimension $1$. Its extension to larger dimensions seems
feasible. Specifically, the forward approximation phase of our algorithm 
will be unproblematic, while in the backward search phase we will have
to give up completeness.

Currently our modeling of a sampled data system has a fixed clock
cycle. It does not seem hard to accommodate variable intervals; such
extension as well as its use is a topic of our future work.

Our modeling benefits a lot from the assumption that the controller
communicates with the plant and the sensor using finite datatypes. 
Some hybrid systems 
do call for relaxation of this assumption in their modeling; it is our
future work to see how the current input synthesis algorithm carries
over
to such relaxation.


\paragraph{Organization of the Paper}
In~\S\ref{section:STAHS} we introduce our modeling of sampled data
systems and formalize the input synthesis
problem. In~\S\ref{section:algorithm} we describe our algorithm,
explaining its three phases one by
one. In~\S\ref{section:optimizationAndImpl} our implementation is
described, together with the experimental results.
The proofs are deferred to the appendix.

\paragraph{Acknowledgments}
We are grateful to the reviewers of an earlier version for their
useful comments and suggestions.
    T.A.\ and I.H.\  are supported by
    Grants-in-Aid for Young Scientists (A) No.\ 24680001,
    and by
    Aihara Innovative Mathematical Modeling Project, FIRST Program,
    JSPS/CSTP;
    K.S. is supported by  Grants-in-Aid for Young Scientists (B) No. 70633692 and
    The Hakubi Project of Kyoto University.

\paragraph{Notations}
$\R$
is the set of real numbers;
$\B = \{\ttrue,\ffalse\}$ is the set of Boolean values.
We let $f[x_0 \mapsto y_0]$ denote \emph{function update}: it carries $x_{0}$ to $y_{0}$ and acts as $f$ on the
    other input.

\section{Modeling Sampled Data Systems}
\label{section:STAHS}

\subsection{Overview}\label{subsection:STAHSIntuition}

\begin{wrapfigure}[9]{r}{0.3\textwidth}
  \begin{equation}\label{equation:structureOfSTAHS}
    \begin{minipage}[c]{.85\textwidth}
      \includegraphics[width=.35\textwidth]{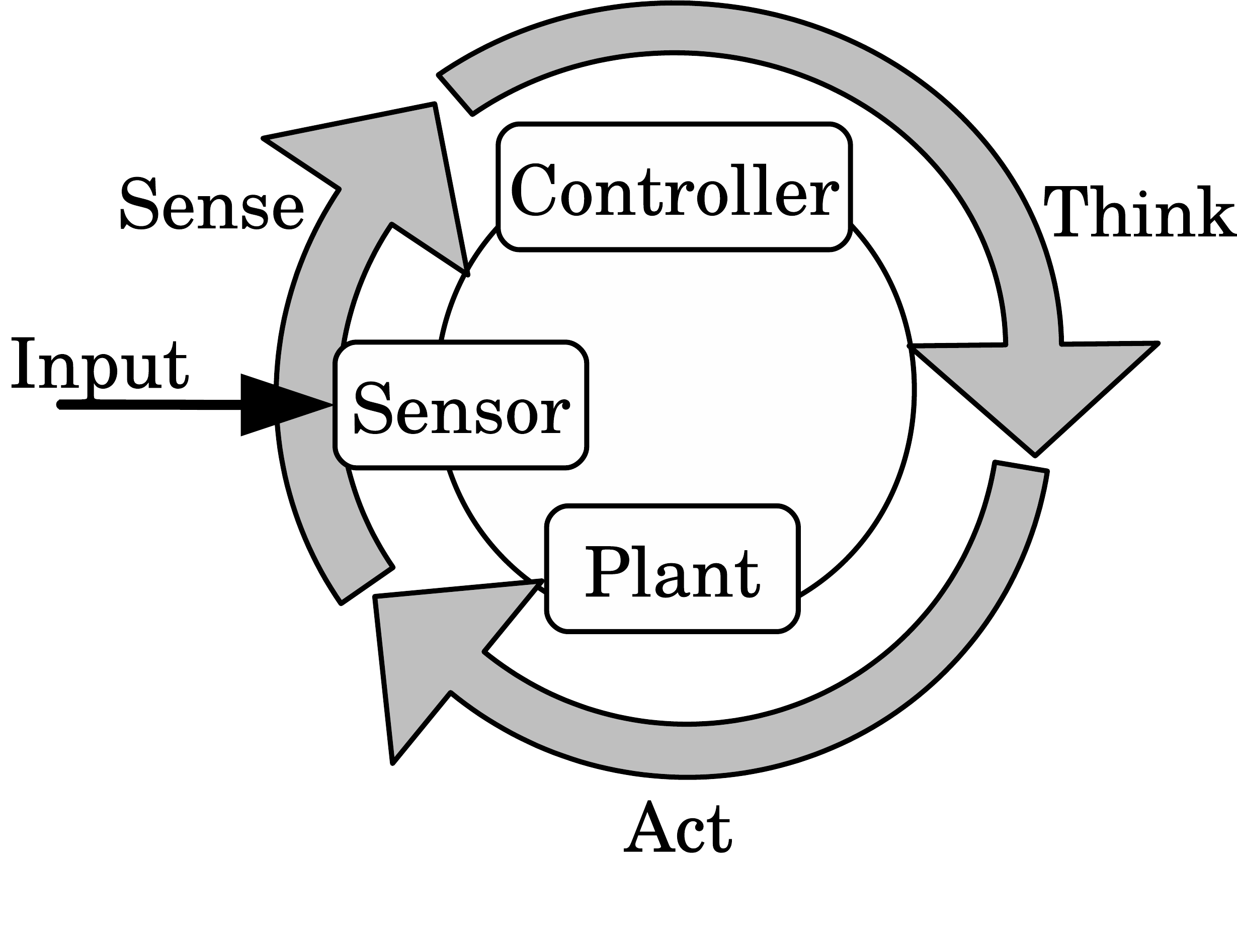}
    \end{minipage}
  \end{equation}
\end{wrapfigure}
\emph{Sampled data systems}  are a class of hybrid systems commonly known
in control theory. In those systems a
physical plant is interrupted by a digital controller in a periodic
manner.
In the current paper where our interests are in input synthesis, it is
convenient to explicitly separate the third component called a
\emph{sensor}. The three components are then organized in a loop,  as
shown on the right in~(\ref{equation:structureOfSTAHS}).



In the execution of sampled data systems thus modeled, we refer to the
three stages in which the sensor, the controller, and the plant
operates, respectively, as the \emph{sense}, \emph{think}, and
\emph{act} stages. Note that the sensor also takes input from outside
the system.


For simplification we further assume the following. 
\begin{enumerate}
 \item A (digital) controller is written in an imperative programming language. 

\item\label{item:fixedInterval} In the execution of a sampled data system, the sense-think-act loop is
  executed at fixed intervals---once every one second. \footnote{The clock cycle can
  be an arbitrary number $\Delta$; in this paper we assume $\Delta=1$ for simplicity.}
\item\label{item:senseAndControlTakesNoTime} The sense and control
  stages take no time for their execution. 
\item\label{item:finitelymanyModes} The controller governs the plant by picking a \emph{mode}, from a finite set
  $\{m_{1},\dotsc,m_{M}\}$.  In particular, the controller cannot
     feed the plant with a continuous value
  $r$.
\item In the act stage the plant operates according to (the ODE
      associated with) the mode $m_{i}$
  picked by the controller. The act stage lasts for one second (a
  fact that follows from~\ref{item:fixedInterval}.\ and \ref{item:senseAndControlTakesNoTime}).
\item\label{item:finitelyManySensorValues} The data sent from the sensor to the controller is finitely many
  Boolean values.
\end{enumerate}


While there are many actual systems that fall out of the realm of this
modeling, it does cover fairly many---among which are fixed interval
digital controllers, a class of hybrid systems ubiquitous in industry.
Sampled data systems, especially under the above assumptions, come to
exhibit pleasant structural properties: its behaviors are much like
those of programs and we can apply forward and backward reasoning in
program logic.  Assumptions~\ref{item:fixedInterval}.\ and
\ref{item:senseAndControlTakesNoTime}.\ are common (see
e.g.~\cite{DBLP:conf/cav/ZutshiST12}). For example, Assumption
\ref{item:senseAndControlTakesNoTime}.\ is reasonable considering the
speed of digital circuits and typical sensing intervals ($\Delta \approx
1 \mathrm{ms}$). Assumptions~\ref{item:finitelymanyModes}.\
and~\ref{item:finitelyManySensorValues}.---that the controller
communicates via finite datatypes---are essential in reducing the input
synthesis problem to a search problem. 


\subsection{The Language $\ImpCtrl$}\label{subsection:loopFreeIMP}
We start with defining an imperative programming language $\ImpCtrl$
that is used to describe the (digital) controller of a sampled data system. It is 
a standard one and is much like $\mathbf{IMP}$ in~\cite{Winskel93}, but
lacks the $\mathtt{while}$ construct. It is indeed unrealistic to 
have $\mathtt{while}$ loops in real-time applications like cyber-physical systems.
Moreover, without $\mathtt{while}$  loops we can succinctly express weakest
preconditions and strongest preconditions---the latter are fully
exploited
in our algorithm for input synthesis.

\begin{wrapfigure}[]{r}{.4\textwidth}
  \includegraphics[width=.4\textwidth]{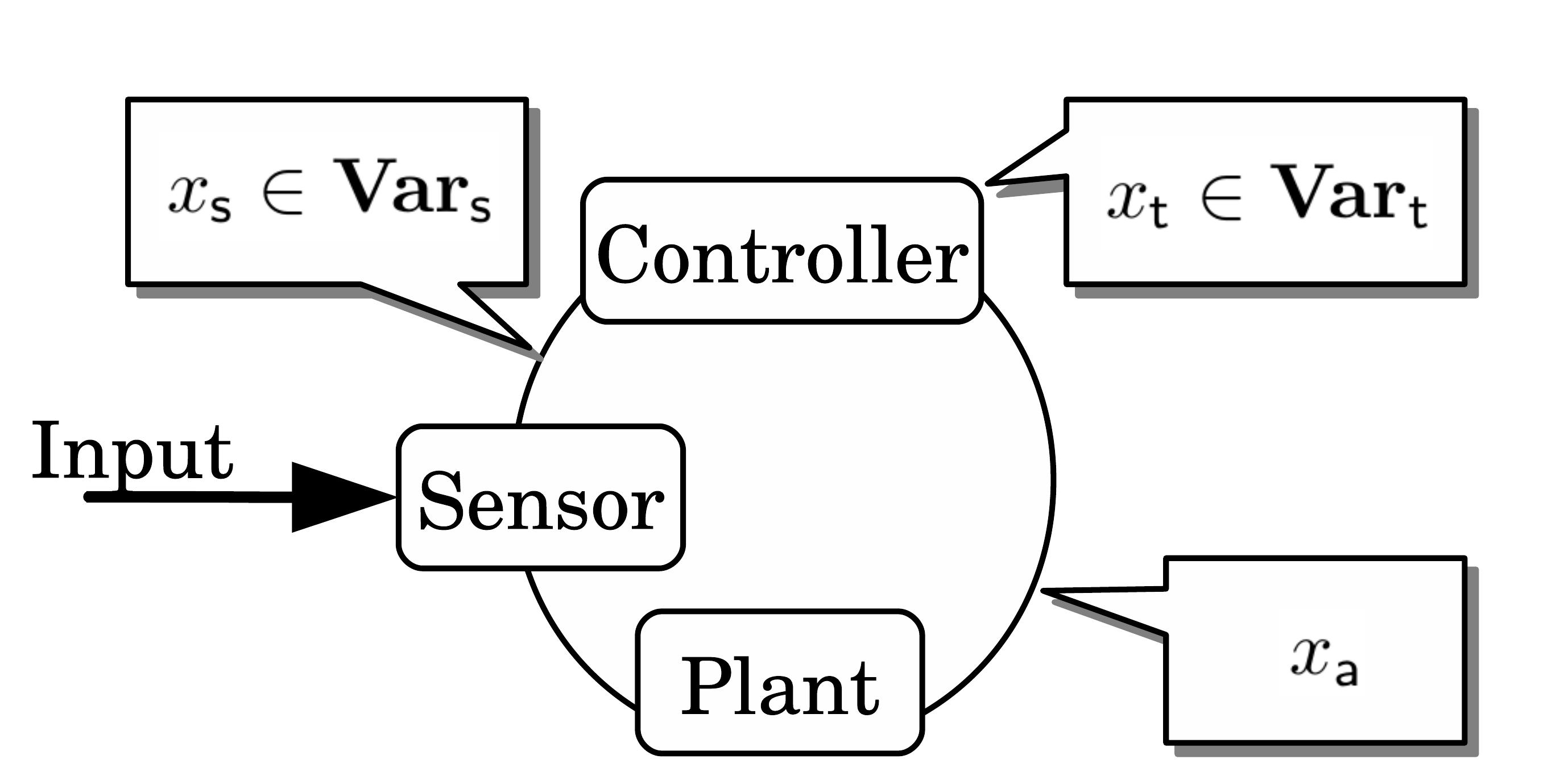}
\end{wrapfigure}
In  $\ImpCtrl$ the set 
$\Var=\Varthink\cup \Varsense \cup \Varact$
of variables is divided into three classes:
the \emph{think}, \emph{sense} and \emph{act} variables. The distinction
is for the purpose of communicating with the other two  components (plant and sensor) of a system. As we will see, a think variable
$x_{\mathsf{t}}\in\Varthink$ stores a real number (which will be a
floating-point number in an actual implementation); a sense variable
$x_{\mathsf{s}}\in\Varsense$ represents a Boolean value sent from the
sensor;
and the (only) act variable $x_{\mathsf{a}}$ in
$\Varact=\{x_{\mathsf{a}}\}$ tells the plant which mode
$m_{i}$ the plant should  take in the coming interval.





\begin{mydefinition}[the language $\ImpCtrl$]
  \label{definition:impCtrl}
  Let $\Modes=\{ m_1, \dotsc, m_{M}\}$ be a fixed finite set of \emph{modes}; 
  $\Varthink$ be a countable set of \emph{think variables};
  $\Varsense$ be a finite set of \emph{sense variables};
  and $\Varact=\{x_\mathsf{a}\}$. The syntax of $\ImpCtrl$ is  as follows.
  \begin{displaymath}
    \begin{array}{rlllll}
      \AExp\ni&
      &a & ::= & r
      \mid \Vthink
      \mid a_1 \aop a_2
      &\text{arithmetic expr.}
      \\
      \BExp\ni&
      &b & ::= & \True
      \mid \False      
      \mid \Vsense
      \mid a_1 \rop a_2 
      \mid \neg b
      \mid b_1 \vee b_2
      \mid b_1 \wedge b_2
      &\text{Boolean expr.}
      \\
      \Cmd\ni&
      &c & ::= & \Skip 
      \mid \Vthink := a 
      \mid \Vact := m_i
      \mid c_1;c_2 
      \mid \If{b}{c_1}{c_2}
      &\text{commands}
    \end{array}
  \end{displaymath}
  Here $r \in \R$, $m_i \in\Modes$, $\Vthink \in \VthinkSet$, $\Vsense
 \in \VsenseSet$, $\aop \in \{+,-,\times\}$ and $\rop\;\in\{=,<
,\le,>,\ge
\}$.
\end{mydefinition}
The semantics of 
 $\ImpCtrl$ is 
as usual, like in~\cite{Winskel93}. See
Def.~\ref{definition:semanticsOfImpCtrl} for details.

\subsection{Assertions for $\ImpCtrl$}\label{subsection:AssnCtrl}
We now introduce an assertion language for $\ImpCtrl$. Its formulas
are used to express pre- and post-conditions in the input synthesis
problem, as well as in program logic. 
The semantics of the first-order language $\AssnCtrl$ is 
as usual. See Def.~\ref{definition:semanticsOfAssnCtrl}.
%
%
\begin{mydefinition}[the assertion language $\AssnCtrl$]
  We fix a set  $\Var'$ of ``logical'' variables such that $\Var \cap \Var' \neq \emptyset$.
  The assertion language $\AssnCtrl$ is defined as follows.
  \begin{displaymath}
    \begin{array}{rllll}
      \AExp\ni
      &a & ::= & r 
      \mid \Vthink
      \mid v'
      \mid a_1 \aop a_2
      &\text{arithmetic expressions}
      \\
      \MExp\ni
      &m & ::= & m_i \mid \Vact
      &\text{mode expressions}
      \\
      \Fml\ni
      &\Phi & ::= & \True
      \mid \False
      \mid \Vsense
      \mid a_1 \rop a_2
      \mid m = m
      \mid \neg \Phi\mid\quad
      &\text{formulas}
      \\
      &&&
      \Phi_1 \vee \Phi_2
      \mid \Phi_1 \wedge \Phi_2
      \mid \forall v' \in \R.\Phi
      \mid \exists v' \in \R.\Phi \\
    \end{array}
  \end{displaymath}
  Here $r \in \R$, $m_i \in \Modes$, $\Vthink \in \VthinkSet$, $\Vsense
  \in \VsenseSet$, and $v' \in \Var'$.
  Intuitively, $\sigma\in\State$ is a valuation that depends on the
 state of a sampled data system; and
  $\gamma\in\R^{\Var'}$ is another valuation of  (logical) variables in $\AssnCtrl$.
\end{mydefinition}


\subsection{Calculi for Weakest Preconditions and Strongest Postconditions}
\label{subsection:weakestPreStrongestPost}

We introduce program logic for $\ImpCtrl$ in the form of 
a \emph{weakest precondition calculus}
(see e.g.~\cite{Winskel93}) and a \emph{strongest
  postcondition calculus} (see e.g.~\cite{GordonC10}). The  calculi will be exploited for the
search space reduction in  input synthesis.

\begin{mydefinition}[weakest precondition $\WPsyntax{c}{\Phi}$;
    strongest postcondition  $\SPsyntax{c}{\Phi}$ ]\label{definition:weakestPreconditionAndStrongestPostcondition}
  Given $c \in \Cmd$ of $\ImpCtrl$ and $\Phi \in \Fml$ of $\AssnCtrl$,
  we define a formula $\WPsyntax{c}{\Phi}\in \Fml$  
  inductively on
  $c$.
  \begin{equation}\label{equation:weakestPrecondCalculus}
    \begin{array}{rlrlrlrl}
      \WPsyntax{\Skip}{\Phi}&\;\equiv\; \Phi \enspace, \quad &
      \WPsyntax{c_1;c_2}{\Phi} &\;\equiv\;
      \WPsyntax{c_1}{\WPsyntax{c_2}{\Phi}} \enspace, &
      \\
      \WPsyntax{\Vthink := a}{\Phi} &\;\equiv\; \Phi[a/\Vthink] \enspace, \quad &
      \WPsyntax{\Vact := m_i}{\Phi} &\;\equiv\; \Phi[m_i/\Vact] \enspace, \\
      \multicolumn{8}{l}{
        \WPsyntax{\If{b}{c_1}{c_2}}{\Phi} \;\equiv\; (b
        \wedge \WPsyntax{c_1}{\Phi}) \vee (\neg b \wedge
        \WPsyntax{c_2}{\Phi})\enspace;
      }
    \end{array}
  \end{equation}
  A formula $\SPsyntax{c}{\Phi}\in\Fml$ is
  defined as follows, similarly by induction.
  \begin{equation}\label{equation:strongestPostcondCalculus}
    \begin{array}{rcl}
      \SPsyntax{\Skip}{\Phi} &\;\equiv\;& \Phi \enspace, \qquad
      \SPsyntax{c_1;c_2}{\Phi} \;\equiv\; \SPsyntax{c_2}{\SPsyntax{c_1}{\Phi}} \enspace, \\
      \SPsyntax{\Vthink := a}{\Phi} &\;\equiv\; & \exists v' \in \R.(\Phi[v'/\Vthink] \wedge \Vthink = a[v'/\Vthink]) \enspace,\\
      \SPsyntax{\Vact := m_i}{\Phi} &\;\equiv\; & (\Phi[m_1/\Vact] \vee \cdots \vee \Phi[m_{M}/\Vact]) \wedge \Vact = m_i \enspace,\\
      \SPsyntax{\If{b}{c_1}{c_2}}{\Phi} &\;\equiv\; & \SPsyntax{c_{1}}{b \wedge \Phi} \vee \SPsyntax{c_{2}}{\neg b \wedge \Phi}\enspace.\\
    \end{array}
  \end{equation}
\end{mydefinition}
In our
implementation,   $\AssnCtrl$ is restricted to its propositional
fragment for  tractability. The quantifier 
in~(\ref{equation:strongestPostcondCalculus}) is thus immediately eliminated
using the quantifier elimination mechanism in Mathematica.
\auxproof{ (like its
 $\mathtt{Resolve}$ function).
}
The third line 
in~(\ref{equation:strongestPostcondCalculus}) is
essentially the same as the second; there we can dispense with a quantifier
$\exists$ since $\Modes=\{ m_1, \dotsc, m_{M}\}$ is a finite set. 

\begin{myproposition}\label{proposition:IndeedWeakestPreconditionAndStrongestPostcondition}
  For any $\sigma \in \State$ and $\gamma \in \R^{\Var'}$,
  \begin{enumerate}
  \item (weakest precondition) $\Assn{\sigma, \gamma}{\WPsyntax{c}{\Phi}}$ if and only if
    $\Assn{\Denote{c}(\sigma), \gamma}{\Phi}$;
  \item (strongest postcondition) $\Assn{\sigma, \gamma}{\Phi}$ if and only if $\Assn{\Denote{c}(\sigma), \gamma}{\SPsyntax{c}{\Phi}}$.
  \end{enumerate}
\end{myproposition}

\subsection{Modeling Sampled Data Systems, Formally}\label{subsection:senseThinkActHybridSys}
We present the formal definition of 
our modeling of sampled data systems, under the assumptions
in~\S{}\ref{subsection:STAHSIntuition}. 


\begin{mydefinition}[sampled data system]
  \label{definition:STAHS}
  Let $n$ be a natural number, and $I\subseteq \R^{n}$ be a fixed set called the
  \emph{input domain}. An \emph{$n$-dimensional sampled data system}  
  is a triple $\Sys = (c,p,s)$ where:
  \begin{itemize}
  \item $c \in \Cmd$ is a command of $\ImpCtrl$
    (\S\ref{subsection:loopFreeIMP}), called a \emph{controller};
  \item $p = \bigl(\, \dot{x} = p_{m_{i}}(t,x) \,\bigr)_{m_{i} \in
    \Modes}$ is a family of (explicit, $n$-dimensional) ODEs indexed by $\Modes=\{ m_1, \dotsc,
    m_{M}\}$, called a \emph{plant}; and
  \item $s: \R^n \times I \to \B^{\VsenseSet}$ is a function, called a {\it sensor}.
  \end{itemize}
  A \emph{state} of a sampled data system is a pair $(\sigma, x)$ 
  of  $\sigma \in
  \State$ and  $x \in \R^n$. In a state $(\sigma,x)$, 
  the component $\sigma$ is called a 
  \emph{controller state(C-state)}, and $x$ a \emph{plant state (P-state)}.
\end{mydefinition}
The dimension $n$ refers to that of the (continuous) plant, meaning that
$x$ and $\dot{x}$ in 
the plant 
$p = (\, \dot{x} = p_{m_{i}}(t,x) \,)_{m_{i} \in
  \Modes}$ 
are vectors in $\R^{n}$.

\begin{myexample}[count and brake]\label{example:CountAndBrake}
  In  Fig.~\ref{fig:ExampleSTA-HS} is a simplification of
  Example~\ref{example:TYTexample};
  this will be our running example.
  The value $v$ is intended to be the velocity of a car.
  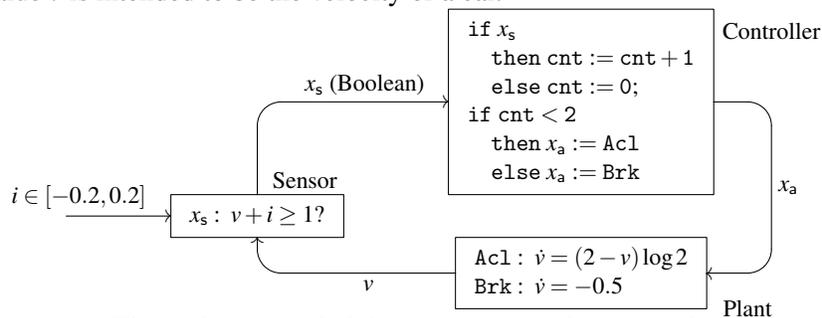
\begin{figure}[htbp]
    \centering
      \scalebox{.9}{
    \begin{math}\small
      \def\labelstyle{\textstyle}
\vcenter{\xymatrix@R=-0em@C+2em{
          & & *+[F]{\begin{array}{l} 
              \tt if\; \mathit{x}_{\mathsf{s}}\\
              \quad \tt then\; cnt := cnt + 1\\
              \quad \tt else\; cnt := 0;\\
              \tt if\; cnt < 2\\
              \quad \tt then\; \mathit{x}_\mathsf{a}:= Acl\\
              \quad \tt else\; \mathit{x}_\mathsf{a}:= Brk
          \end{array}}
          \shifted{8em}{3em}{\text{Controller}}
          \ar`r[dd]+/r8em/`[dd]^{x_\mathsf{a}}[dd]
          \\
            {}
            \ar[r]^*[r]{i\in[-0.2,0.2]\quad}
            &
            *+[F]{
              \begin{array}{l} 
                \Vsense:\; v+i\ge 1 ?
            \end{array}}
            \shifted{2em}{1.5em}{\text{Sensor}}   
            \ar`u[ru] [ru]^-{\Vsense \text{ (Boolean)}}
            \\
            & & *+[F]{\begin{array}{l} 
                {\tt Acl}:\; \dot{v}=(2-v)\log{2}
                \\
                  {\tt Brk}:\; \dot{v}=-0.5
            \end{array}}
            \shifted{7em}{-1.5em}{\text{Plant}}
            \ar`l[lu]^-{v} [lu]
      }}
    \end{math}
   }
    \abovecaptionskip = 0pt
    \caption{A sampled data system (running example)}
    \label{fig:ExampleSTA-HS}
  \end{figure}%

  The example follows a pattern of fixed interval controllers commonly
  used in industry. Namely, a counter $\mathtt{cnt}$ is used to tell if 
  extremity ($v+i\ge 1$) has continued for a certain critical
  number of intervals ($2$ here). If $\mathtt{cnt}$ reaches the critical
  number a countermeasure is taken: the plant is set to the braking mode
  ($\mathtt{Brk}$) and the velocity $v$ decreases.  Otherwise the plant
  operates in the acceleration mode ($\mathtt{Acl}$), which is a first-order
  lag system where the velocity $v$ approaches towards $2$.

  The system takes input $i$---whose domain is assumed to be $[-0.2,0.2]$---that models disturbance from outside. For
  example, the road can be slippery, which can make the actual velocity $v$
  different from the value that is used by the controller.

\end{myexample}

\subsection{Semantics of Sampled Data Systems}
\label{subsection:semanticsOfSTAHS} 
We formally define the semantics of
a sampled data system. Our current concern is not so much on the
solution of ODEs as on the interaction between a controller and a
plant. Therefore we adopt the following black-box view of a plant.
\begin{mydefinition}[$\Execplant(p,x)$]\label{definition:execPlant}
  In what follows we assume that all the ODEs used for a plant have unique solutions. That is, for any $n$-dimensional ODE
  $\dot{x}=p(t,x)$ and an initial value $x_{0}\in\R^{n}$, we assume that there exists
  a unique function $F:[0,1]\to \R^{n}$ such that: $F(0)=x_{0}$; and
  for any $t\in[0,1]$, $\dot{F}(t)=p(t,F(t))$.

  By $\Execplant(p,x_{0})$ we denote the state of the plant
  $\dot{x}=p(t,x)$ at time $t=1$, assuming that the initial state 
  (at time $t=0$) is $x_{0}$. That is, $\Execplant(p,x_{0})=F(1)$ where
  $F$
  is the function in the above.
\end{mydefinition}

In our implementation we actually use the result of numerical calculations (by
MATLAB) as the value $\Execplant(p,x)$, ignoring numerical errors.


\begin{mydefinition}[semantics of a sampled data systems]\label{definition:semanticsOfSTAHS}
  Let $\Sys = (c,p,s)$ be a sampled data system. The \emph{one-step transition}
  is a ternary relation $\to$ among two states $(\sigma,x)$, $(\sigma',x')$
  and input $i\in I$; this is denoted by
  $(\sigma,x)\xrightarrow{i}(\sigma',x')$. It is defined as follows.

  We have   $(\sigma,x) \xrightarrow{i} (\sigma',x')$ if
  $(\sigma', x') = (\Act_{\Sys} \circ \Think_{\Sys} \circ
  \Sense_{\Sys})(\sigma, x, i)$, where the three functions are defined by:
  \begin{equation}\label{equation:semanticsOfSTAHS}
    \begin{array}{rlll}
      \Sense_{\Sys}\;:\; & \State \times X \times I  \longrightarrow
      \State \times X\enspace,\quad
      & (\sigma, x, i)  \longmapsto  \bigl(\,\sigma[\Vsense \mapsto s(x,i)(\Vsense)], x\,\bigr)\enspace;\\
      \Think_{\Sys}\;:\; & \State \times X  \longrightarrow \State \times X\enspace,\quad
      & (\sigma, x)  \longmapsto  \bigl(\Denote{c}(\sigma), x\bigr)\enspace;\\
      \Act_{\Sys}\;:\; & \State \times X \longrightarrow  \State \times X\enspace,\quad
      & (\sigma, x)  \longmapsto  \bigl(\sigma,
      \Execplant(p_{\sigma(\Vact)},x)\bigr) 
      \enspace.\\
    \end{array}
  \end{equation}
  Here $\Denote{c}$ is as in Def.~\ref{definition:semanticsOfImpCtrl}.
  It is clear that, given a state $(\sigma,x)$ and $i\in I$, the
  post-state $(\sigma',x')$ such that $(\sigma,x) \xrightarrow{i}
  (\sigma',x')$ is uniquely determined.
  A succession 
  $(\sigma_0,x_0) \xrightarrow{i_0} (\sigma_1,x_1) \xrightarrow{i_1} \cdots \xrightarrow{i_{T-1}} (\sigma_T,x_T)$ 
  of one-step transition is called a \emph{run} of the system $\Sys$.

\end{mydefinition}

A specification of a state of a sampled data system is given by a pair of an
assertion formula (on the controller) and a subset of $\R^{n}$ (on the
plant).
%
%


\begin{mydefinition}[CP-condition]\label{definition:conditionOnSTAHS}
  Let $\Sys = (c,p,s)$ be an $n$-dimensional sampled data system.
  A \emph{controller-plant condition (CP-condition)} for $\Sys$ is 
  a pair $(\Phi,X)$ of 
  an assertion $\Phi \in \Fml$ called the \emph{controller condition} 
  and a condition $X \subseteq \R^n$ called the \emph{plant condition}.
  The projection to each component 
  is denoted by $\CtrlCond$ and $\PlantCond$ respectively.

  Given a state $(\sigma, x) \in \State \times \R^n$ of $\Sys$ and a
  CP-condition $(\Phi, X)$, we write 
  $\Assn{(\sigma,x)}{(\Phi, X)} $ if
  $\Assn{\sigma}{\Phi}$ and
  $x \in X$.
  $(\Phi, X)$ is \emph{satisfiable} if there is a state
  that satisfies it.
\end{mydefinition}

\subsection{The Input Synthesis Problem for Sampled Data Systems}
\label{subsection:inputSynthesis}


\begin{mydefinition}[input synthesis problem]\label{definition:inputSynthesisProblem}
  The \emph{input synthesis problem} 
 is:
  \begin{center}
    \begin{tabular}{ll}
      given:&
      \begin{minipage}[t]{.9\textwidth}
        \begin{itemize}
        \item $\Sys = (c,p,s)$,  an $n$-dimensional sampled data system;
        \item $(\Phi_{\init}, X_{\init})$ and $(\Phi_{\final}, X_{\final})$, a pre- and a post-CP-condition; and
        \item $T \in \N$, the number of steps,
        \end{itemize}
      \end{minipage}
      \\
      return:&
      \begin{minipage}[t]{.9\textwidth}
        \begin{itemize}
        \item  
          an initial state $(\sigma_0, x_0) \in \State \times \R^{n}$
	  such that
          $\Assn{(\sigma_0, x)}{(\Phi_{\init}, X_{\init})}$;
          and
        \item an input sequence $i_0, \dotsc, i_{T-1} \in I$ such that, 
	  for the corresponding run
          $(\sigma_0,x_0) 
          \xrightarrow{i_0} (\sigma_1,x_1) 
          \xrightarrow{i_1}
	  \cdots 
           \xrightarrow{i_{T-1}}
          (\sigma_T,x_T)$ of $\Sys$, we
	  have $\Assn{(\sigma_T, x_T)}{(\Phi_{\final}, X_{\final})}$.
        \end{itemize}
      \end{minipage}
    \end{tabular}
  \end{center}
\end{mydefinition}

\begin{myexample}\label{example:leadingExampleProblem}
  Let $\Sys$ be the sampled data system in
  Example~\ref{example:CountAndBrake}.  Consider 
  \begin{displaymath}
    \text{ a
      pre-CP-condition}
    \quad
    (\cnt = 0, \enspace [0, 1])
    \quad
    \text{and a
      post-CP-condition}
    \quad
    (\True, \enspace [1.5, 2])
  \end{displaymath}
  and $T=4$ as the number of steps. 
\auxproof{The number $T$ is kept small since
  the current example will be used in demonstrating our input synthesis
  algorithm. 
}  
  In the input synthesis problem, we seek for an initial state
  $(\sigma_0, x_0)$  and an input sequence 
  $i_0, i_{1},i_{2}, i_{3} \in [-0.2, 0.2]$ such that
  \begin{displaymath}
    \begin{array}{l}
      (\sigma_0, x_0) \models (\cnt = 0, \enspace [0, 1]) \enspace,
      \;
      (\sigma_0,x_0) \xrightarrow{i_0} (\sigma_1,x_1) \xrightarrow{i_1} \cdots
      \xrightarrow{i_3} (\sigma_4,x_4)
      \; \text{and} \; (\sigma_4, x_4) \models (\True, \enspace [1.5, 2])\enspace.
    \end{array}
  \end{displaymath}
\end{myexample}

\section{An Algorithm for Input Synthesis for Sampled Data Systems}
\label{section:algorithm}
In this section we present our algorithm. We identify the core of the
input synthesis problem to be
the discovery of suitable
  input and output of the controller 
at each step. More specifically, we seek for
a \emph{successful path}
\begin{equation}\label{equation:theSequenceWeLookFor}
  \overrightarrow{(\sigma_{\mathsf{s}}, m)}
  \;:=\;
  \bigl\langle
  \;
  (\sigma_{\mathsf{s}}^{(T-1)}, m^{(T-1)})
,\,
  (\sigma_{\mathsf{s}}^{(T-2)}, m^{(T-2)})
,\,
  \dotsc,\,
 (\sigma_{\mathsf{s}}^{(0)}, m^{(0)})
  \;\bigr\rangle
\end{equation}
where $\sigma_{\mathsf{s}}^{(k)}:\VsenseSet\to \B$ is a valuation of
sense variables---which shall be henceforth called \emph{sensor
output}---and $m^{(k)}\in \Modes$ is a mode.\footnote{Note that time is
reversed in~(\ref{equation:theSequenceWeLookFor}). This is purely for
the purpose of presentation.}  Together with an initial state
$(\sigma_{0},x_{0})$, the sensor output $\sigma^{(k)}_{\mathsf{s}}$
determines the behavior of the controller, and the mode $m^{(k)}$
determines that of the plant, at each step $k$.  Therefore a path like
in~(\ref{equation:theSequenceWeLookFor}) determines the behavior of the
whole sampled data system from step $0$ through step $T$; a
``successful'' path is then one that steers the given precondition to
the given postcondition.

Towards the discovery of a successful 
path,
our approach is to  exploit the program logic
in~\S{}\ref{subsection:weakestPreStrongestPost}---i.e.\ to make  most
of the structure of
the controller as a program.  
In our modeling of sampled data systems (\S{}\ref{section:STAHS}) we
have made assumptions so that the program-logic approach is possible.


Concretely, our algorithm consists of the following three
phases.
\begin{enumerate}
\item\textbf{(Forward approximation)} 
  We  overapproximate the set of CP-states that the system can 
  reach, starting from the pre-CP-condition $(\Phi_{\init}, X_{\init})$
  and going forward step by step. This first phase is seen as a preparation
  for the second (main) phase.



\item \textbf{(Backward search)} 
  A successful path~(\ref{equation:theSequenceWeLookFor}) will be a path
  in a so-called \emph{backward search tree}. Its
   branching degree
  is $2^{|\VsenseSet|}\times |\Modes|$; its nodes are labeled with
  CP-conditions; and its root is labeled with the post-CP-condition
  $(\Phi_{\final}, X_{\final})$. We search for a successful path in the
      tree, in a depth-first manner.

\item \textbf{(Synthesis of actual input)} 
  We choose an initial state $(\sigma,x)$ and go on to  synthesize an
  input sequence $i_{0},\dotsc, i_{T-1}$,
  using the successful path $\overrightarrow{(\sigma_{\mathsf{s}}, m)}$
  discovered in the  previous phase. 
  This can be done in a straightforward linear manner.
\end{enumerate}
The second phase (backward search) is where
an actual (depth-first) search is done. Program logic
is used there to prune branches
and reduce the search space. 



\subsection{Forward Approximation}
\label{subsection:fwdApprox}
In this phase of the algorithm,
we overapproximate the behavior of the given sampled data system and obtain a
sequence 
$\bigl( \kFA(\Phi_{\init},X_{\init}) \bigr)_{0 \leq k \leq T}$ 
of CP-conditions.
These are
obtained iteratively as follows.


\begin{mynotation}[$s^{-1}$]
  \label{notation:sInverse}
  Let 
  $\Sys = (c,p,s)$ be an $n$-dimensional sampled data system; $I$ be its
  input domain; and $\sigma_{\mathsf{s}}\in\B^{\VsenseSet}$ be 
  sensor output. We abuse notation and  denote  by $s^{-1}(\sigma_{\mathsf{s}})$  the set of plant states 
  that can be ``steered'' to $\sigma_{\mathsf{s}}$. Precisely,
  \begin{math}
    s^{-1}(\sigma_{\mathsf{s}})
    \;:=\;
    \{ x \in \R^{n} \mid \exists i \in I.\,s(x,i) = \sigma_s \}\enspace.
  \end{math}
\end{mynotation}
For example, let $\sigma_{\mathsf{s}}$ such that
$\sigma_{\mathsf{s}}(\Vsense)=\ttrue$
in the setting of Example~\ref{example:CountAndBrake}.
We have
\begin{math}
  s^{-1}(\sigma_{\mathsf{s}})  = \{ x \in X \mid \exists i \in [-0.2, 0.2].\, x + i \geq 1 \}
  = [0.8, \infty)
\end{math}.
\auxproof{ Similarly, for $\sigma_{\mathsf{s}}'$ such that $\sigma'_{\mathsf{s}}(\Vsense) = \ffalse$,  we have $s^{-1}(\sigma_{\mathsf{s}}') = (-\infty, 1.2)$.
}


\begin{mydefinition}[$\OneFA, \kFA$]
  \label{definition:oneFAkFA}
  Let  $\Sys = (c,p,s)$ be a
sampled data system.
  Let us first define the functions $\PreOneFA_{\Sense}$,
  $\PreOneFA_{\Think}$ and $\PreOneFA_{\Act}$ as follows. Their types
  should be obvious.
  \begin{equation}\label{equation:defOfStagesOfOneFA}
    \begin{array}{rl}
      \PreOneFA_{\Sense}(\sigma_{\mathsf{s}})(\Phi,X) 
      &\;:=\;
      \bigl(\,\SPsyntax{x_{\mathsf{s}} := \sigma_{\mathsf{s}}(x_{\mathsf{s}})}{\Phi},\, X \cap s^{-1}(\sigma_{\mathsf{s}})\,\bigr)\\
      \PreOneFA_{\Think}(\Phi,X)
      &\;:=\;
      \bigl(\,\SPsyntax{c}{\Phi},\, X\,\bigr)\enspace,
      \\
      \PreOneFA_{\Act}(m)(\Phi,X)
      &\;:=\;
      \bigl(\,\Phi \wedge x_{\mathsf{a}} = m,\, \Execplant(p_m, X)\,\bigr)\enspace.
    \end{array}
  \end{equation}
  Here $\SPsyntax{c}{\Phi}$ in the second line is the strongest postcondition
  (Def.~\ref{definition:weakestPreconditionAndStrongestPostcondition}); 
  $\Execplant(p_m, X)$ in the third line is the direct image of
  $X\subseteq \R^{n}$ by the function in Def.~\ref{definition:execPlant};
  and $\SPsyntax{x_{\mathsf{s}} :=
    \sigma_{\mathsf{s}}(x_{\mathsf{s}})}{\Phi}$
  in the first line is defined as follows, similarly to 
  Def.~\ref{definition:weakestPreconditionAndStrongestPostcondition}.
  \begin{displaymath}
    \SPsyntax{x_{\mathsf{s}} :=
      \sigma_{\mathsf{s}}(x_{\mathsf{s}})}{\Phi} 
    \;:\equiv\;
    \begin{cases}
      (\Phi[\True/\Vsense] \vee \Phi[\False/\Vsense]) \wedge \Vsense & \mathrm{if}\; \sigma_{\mathsf{s}}(\Vsense) = \True\\
      (\Phi[\True/\Vsense] \vee \Phi[\False/\Vsense]) \wedge \neg \Vsense & \mathrm{if}\; \sigma_{\mathsf{s}}(\Vsense) = \False\\
    \end{cases} 
  \end{displaymath}
  These three functions are composed to yield:
  \begin{displaymath}
    \PreOneFA(\sigma_{\mathsf{s}},m)(\Phi,X) 
    \;:=\; \PreOneFA_{\Act}(m)\bigl( \PreOneFA_{\Think} \bigl( \PreOneFA_{\Sense}(\sigma_{\mathsf{s}})(\Phi,X) \bigr) \bigr)\enspace;
  \end{displaymath}
  this is understood as the strongest postcondition after the one-step execution of 
  $\Sys$, \emph{assuming that the sensor output $\sigma_{\mathsf{s}}$
    and the mode $m$ have been chosen.}

  Finally, the \emph{one-step forward approximation} function is defined 
  as the following disjunction/union over different 
  $\sigma_{\mathsf{s}}$ 
  and  $m$:
  \begin{equation}\label{equation:oneStepFwdApprox}
\begin{array}{l}
     \OneFA(\Phi,X) :=
    \biggl(\; \bigvee_{(\sigma_{\mathsf{s}}, m) \in \mathcal{M}} \CtrlCond \bigl(\PreOneFA(\sigma_{\mathsf{s}},m)(\Phi,X) \bigr), 
    \bigcup_{(\sigma_{\mathsf{s}}, m) \in \mathcal{M}} \PlantCond
    \bigr(\PreOneFA(\sigma_{\mathsf{s}},m)(\Phi,X) \bigr)
    \;\biggr)\enspace,
 \\
 \qquad\qquad\text{where}\quad\mathcal{M} := \{\, (\sigma, m) \in \B^\VsenseSet \times \Modes
  \mid \PreOneFA(\sigma_{\mathsf{s}}, m)(\Phi,X) \text{ is satisfiable.}
  \}
\end{array}  
\end{equation}
 The projections $\CtrlCond$ and $\PlantCond$, as well as
satisfiability of CP-conditions, are
from
  Def.~\ref{definition:conditionOnSTAHS}.


  We write $\kFA(\Phi, X)$ for $(\OneFA)^{k}(\Phi, X)$. The sequence
  $\bigl( \kFA(\Phi_{\init},X_{\init}) \bigr)_{0 \leq k \leq T}$ of
  CP-conditions  is called the \emph{forward approximation
    sequence} for $\Sys$.

\end{mydefinition}


As an example we present forward approximation for
Example~\ref{example:leadingExampleProblem}. The first one-step
approximation (from $k=0$ to $1$)
is shown below, stage by stage. 
\begin{equation}
  \label{equation:fwdApproxStageByStage}\scriptsize
\scalebox{.9}{\begin{math}
   \vcenter{\xymatrix@C=1.8em@R=-1.9em{
      {k=0}
      \ar@{=>}[r]^*+{\text{\sf sense}}
      &
          {}
          \ar@{=>}[r]^*+{\text{\sf think}}
          &
              {}
              \ar@{=>}[r]^*+{\text{\sf act}}
              &
                  {}
                  \ar@{=>}[r]^*+{\text{\sf unify}}
                  &
                      {k=1}
                      \\
                        {\phantom{
                            \begin{array}{c}
                              hoge\\hoge\\hoge
                            \end{array}
                        }}
                        \\
                        &&&
                        *+[F]{
                          \begin{array}{c} 
                            \cnt = 1 
                            \\
                            \land \Vact={\tt Acl}\land \Vsense
                            \\
                              {[1.4 , 1.5]}
                        \end{array}}
                        \ar[rddd]
                        \\
                        &
                        *+[F]{
                          \begin{array}{c} 
                            \cnt = 0 
                            \land \Vsense
                            \\
                              {[0.8 , 1]}
                        \end{array}}
                        \ar[r]
                        &
                        *+[F]{
                          \begin{array}{c} 
                            \cnt = 1
                            \\
                            \land \Vact={\tt Acl}
                            \land \Vsense
                            \\
                              {[0.8 , 1]}
                        \end{array}}
                        \ar[ru]^*+{\tt Acl}
                        \ar[rd]_*+{\tt Brk}
                        \\
                        &&&
                        *+[F]{
                          \begin{array}{c} 
                            \False
                            \\
                              {[0.3 , 0.5]}
                        \end{array}}
                        \\
                        *+[F]{
                          \begin{array}{c} 
                            \cnt = 0
                            \\
                              {[0,1]}
                        \end{array}}
                        \ar[ruu]^(.4)*+{\Vsense \mapsto \ttrue}
                        \ar[rdd]_(.4)*+{\Vsense \mapsto \ffalse}
                        &&&&
                        *+[F]{
                          \begin{array}{c} 
                            (\cnt = 0\lor\cnt = 1) 
                            \\\land \Vact={\tt Acl}
                            \\
                              {[1,1.5]}
                        \end{array}}
                        \\
                        &&&
                        *+[F]{
                          \begin{array}{c} 
                            \cnt = 0
                            \\
                            \land \Vact={\tt Acl}
                            \land \lnot \Vsense
                            \\
                              {[1,1.5]}
                        \end{array}}
                        \ar[ru]
                        \\
                        &
                        *+[F]{
                          \begin{array}{c} 
                            \cnt = 0
                            \land \lnot \Vsense
                            \\
                              {[0,1]}
                        \end{array}}
                        \ar[r]
                        &
                        *+[F]{
                          \begin{array}{c} 
                            \cnt = 0
                            \\
                            \land \Vact={\tt Acl}
                            \land \lnot \Vsense
                            \\
                              {[0,1]}
                        \end{array}}
                        \ar[ru]^*+{\tt \;\; Acl}
                        \ar[rd]_*+{\tt \;\; Brk}
                        \\
                        &&&
                        *+[F]{
                          \begin{array}{c} 
                            \False
                            \\
                              {[-0.5,0.5]}
                        \end{array}}
  }}
\end{math}
}
\end{equation}
Observe that we
have
four CP-conditions in the fourth column from the left.
Each of them corresponds to a choice of
$(\sigma_{\mathsf{s}}, m)$. Two among the four CP-conditions are
unsatisfiable and hence discarded (i.e.\ they are not in $\mathcal{M}$); the remaining two are unified
and yield $\OneFA(\cnt=0,[0,1])$ in the rightmost column.
\footnote{
  Our approximation can be finer:
  in~(\ref{equation:fwdApproxStageByStage}), in the unification stage, the correlation between a C-condition and a
  P-condition is forgotten by separately taking the disjunction of
  C-conditions
  and the union of P-conditions
  (see~(\ref{equation:oneStepFwdApprox})). Finer approximation, however, 
  makes the approximants grow much bigger and slows down 
  the  backward search phase of the algorithm.
}

\begin{wrapfigure}[11]{r}{.5\textwidth}
  \begin{equation}
    \label{equation:fwdApproxPicture}
    \includegraphics[width=.47\textwidth]{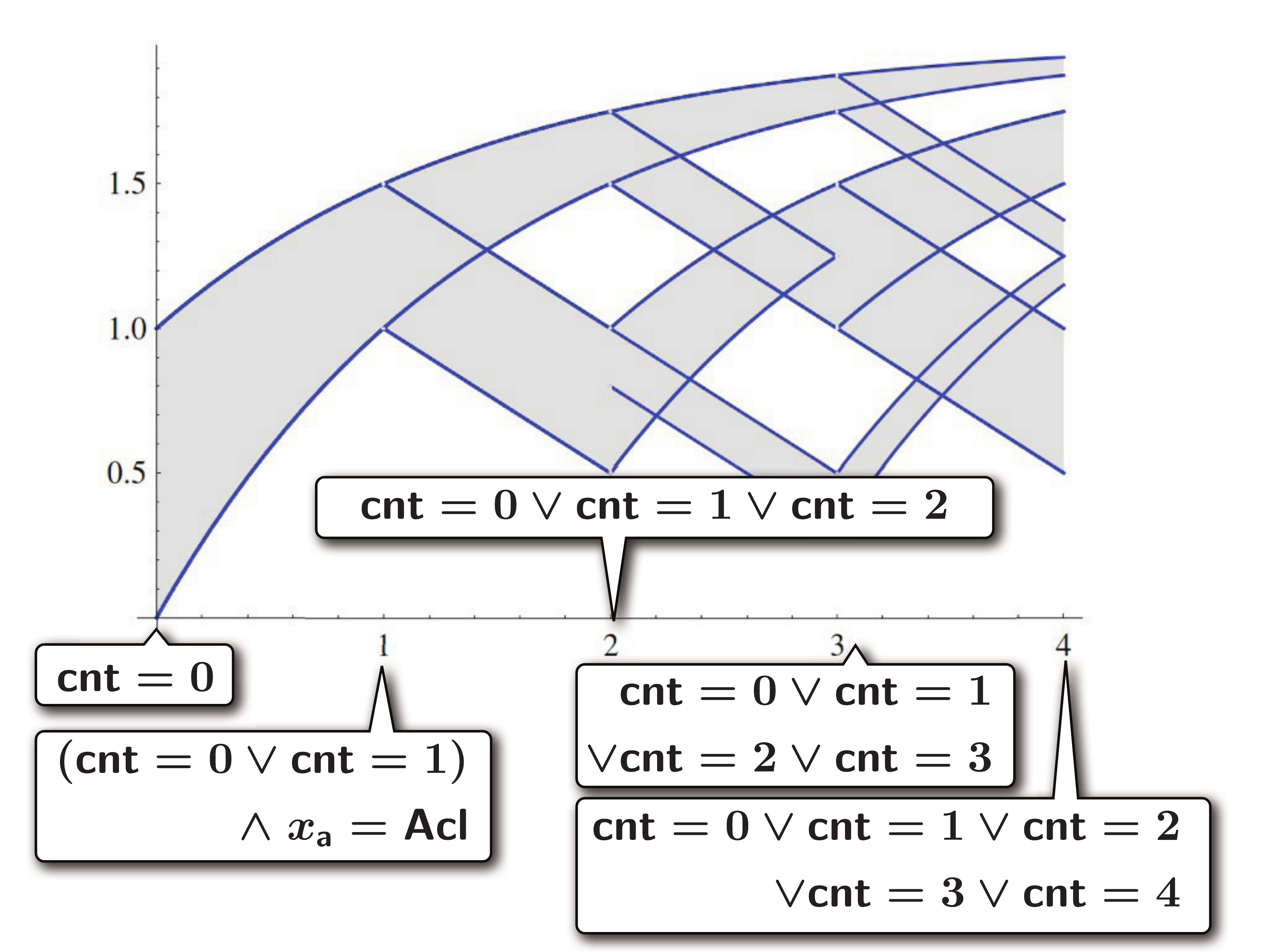}
  \end{equation}
\end{wrapfigure}

By continuing further
we obtain the  forward approximation sequence 
shown on the below
 in~(\ref{equation:fwdApproxPicture}), presented pictorially.

For the completeness of our algorithm
we need to prove that our forward
approximation is indeed  an
\emph{over}-approximation.

%

\begin{myproposition}\label{proposition:1stepForwardApproxCorrectness}
  Let $i_{0},\dotsc,i_{k-1}\in I$ be any input sequence;
  $(\sigma,x)
  \xrightarrow{i_0} \cdots \xrightarrow{i_{k-1}} (\sigma',x')$
  be a run of $\Sys$;
  and $(\sigma,x)\models (\Phi,X)$. Then 
  \begin{math}
    (\sigma',x')
    \models
    \kFA(\Phi,X)
  \end{math}.
  \myqed
\end{myproposition}


\subsection{Backward Search}
\label{subsection:backwardSearch}
In this phase of the algorithm we search for a successful path
$\overrightarrow{(\sigma_{\mathsf{s}}, m)}$ of sensor output and
modes---i.e.\ one that  steers an initial state to a desired postcondition.
The search is conducted in a backward depth-first manner in a tree
called the \emph{backward search tree}.


For  the input synthesis problem,
it is not necessary to construct the whole backward search tree:
finding a leaf whose CP-condition is compatible with the precondition suffices. We will use 
program logic
(\S{}\ref{subsection:weakestPreStrongestPost})---and the forward
approximation sequence obtained in the previous phase---in
pruning branches and reducing the search space. 

\begin{mydefinition}[backward search tree]\label{definition:backwardSearchTree}
  Given an input synthesis problem, its \emph{backward search tree} is a
  tree with branching degree $2^{|\VsenseSet|}\times |\Modes|$
  and with height $T + 1$.  The nodes of the tree are defined inductively
  as follows.
  \begin{itemize}
  \item The root of the tree is labeled with  the postcondition $(\Phi_{\final}, X_{\final})$.
  \item Let $(\Phi,X)$ be the label at the position
    \begin{math}
      \overrightarrow{(\sigma_{\mathsf{s}},m)}
      =
      \bigl\langle\,
      (\sigma_{\mathsf{s}}^{(T-1)},m^{(T-1)}),\,
      \dotsc,\,
      (\sigma_{\mathsf{s}}^{(k+1)},m^{(k+1)})
      \,\bigr\rangle
    \end{math}.
    Its child at the position
    \begin{math}
      \overrightarrow{(\sigma_{\mathsf{s}},m)}
      (\sigma_{\mathsf{s}}',m')
    \end{math}
    is labeled by
    \begin{equation}\label{equation:childInBackwardSearchTree}
      \begin{array}{lll}
        (\Phi',X') \,:=\,&
        \Bigl(\,& \CtrlCond \bigl( \kFA(\Phi_{\init}, X_{\init}) \bigr) \wedge \CtrlCond(\PreBS(\sigma_{\mathsf{s}}, m)(\Phi, X)),\\
        &     & \PlantCond \bigl( \kFA(\Phi_{\init}, X_{\init}) \bigr) \cap \PlantCond(\PreBS(\sigma_{\mathsf{s}}, m)(\Phi, X)) \;\bigr)\enspace,\\
      \end{array}
    \end{equation}
    where the function $\PreBS$ is defined as follows.
    \begin{equation}\label{equation:preBackwardSearch}
      \begin{array}{rll}
        \PreBS(\sigma_{\mathsf{s}}, m)(\Phi, X) &\;:=\;&
	 \PreBS_{\Sense}(\sigma_{\mathsf{s}})\Bigl( \PreBS_{\Think}
	 \bigl( \PreBS_{\Act}(m)(\Phi,X) \bigr) \Bigr)
        \;\text{where}
        \\
        \PreBS_{\Act}(m)(\Phi,X) &\;:=\;& \bigl(\Phi \wedge x_{\mathsf{a}} = m, \Execplant(\rev(p_{m}), X)\bigr)\enspace,\\
        \PreBS_{\Think}(\Phi,X) &\;:=\;& (\WPsyntax{c}{\Phi},
	 X)\enspace,
	 \quad\text{and}\\
        \PreBS_{\Sense}(\sigma_{\mathsf{s}})(\Phi,X) &\;:=\;&
	(\Phi[\sigma_{\mathsf{s}}(\Vsense)/\Vsense], X \cap
	s^{-1}(\sigma_{\mathsf{s}}))\enspace.
      \end{array}
    \end{equation}
    In the second line, $\Execplant(\rev(p_{m}),x_{1})$ means running the 
    original ODE $\dot{x} = p_{m}(t,x)$ with time reversed (i.e.\ from
    $t=1$ to $t=0$) and with the ``initial'' value $x_{1}$ (at time
    $t=1$). Concretely,
    $\rev(p_{m})$ is given by: $\rev(p_{m})(t, x)=-p_{m}(1-t,x)$.
  \end{itemize}

  By 
  $  \BS
  \bigl(
  \overrightarrow{(\sigma_{\mathsf{s}},m)}
  \bigr)$ we denote the label in the tree, at the position designated by
  the path $\overrightarrow{(\sigma_{\mathsf{s}},m)}$. 
  $  \BS
  \bigl(
  \overrightarrow{(\sigma_{\mathsf{s}},m)}
  \bigr)$ 
  is therefore
  a CP-condition.

\end{mydefinition}

\begin{mydefinition}[successful path]\label{definition:successfulPath}
  Let 
  \begin{math}
    \overrightarrow{(\sigma_{\mathsf{s}},m)}
  \end{math}
  be a path in the backward search tree.
  It is \emph{successful}
  if: 1) it is of length $T$; and 2) the label
  $ \BS
  \bigl(
  \overrightarrow{(\sigma_{\mathsf{s}},m)}
  \bigr)$
  at the leaf
  is satisfiable.
\end{mydefinition}


The following
establishes that
that
finding a successful path 
in the backward search tree
 is equivalent to 
solving the input synthesis problem.


\begin{myproposition}[soundness \& completeness]\label{proposition:correctnessOfBwdSearchTree}
  Let 
  \begin{math}
    \overrightarrow{(\sigma_{\mathsf{s}},m)}
  \end{math}
  be a successful path in the backward search tree. 
  Assume also that $\sigma_{0} \in \State$ and $x_{0} \in \R^n$ satisfy $(\sigma_{0}, x_{0}) \,\models\,
  \BS
  \bigl(
  \overrightarrow{(\sigma_{\mathsf{s}},m)}
  \bigr)
  $. Then
  there exists an input sequence $i_{0},\dotsc, i_{T-1}\in T$
  such that an initial state $(\sigma_{0}, x_{0})$ together with $i_{0},\dotsc, i_{T-1}$ is an answer to the input synthesis problem.

  Conversely,   assume  there is
  an answer to an input synthesis problem, given by
  $(\sigma_0, x_0)$ and $i_{0},\dotsc, i_{T-1}$.
  Then there is a successful path
  $\overrightarrow{(\sigma_{\mathsf{s}},m)}$. 
  \myqed  
\end{myproposition}



In searching for a successful path in a backward search tree,
once 
we hit an unsatisfiable label,
clearly all its offspring are unsatisfiable.
We therefore prune such a branch.  The use of $\kFA$
 in~(\ref{equation:childInBackwardSearchTree}) strengthens the labels
 and  makes more branches pruned.

\begin{mylemma}[pruning is correct]\label{lemma:pruningIsCorrect}
  In the backward search tree, 
  assume that the label at the position $\overrightarrow{(\sigma_{\mathsf{s}},m)}$ is unsatisfiable.
  Then its child
  has an unsatisfiable label too.
  \myqed
\end{mylemma}

An example is again using
Example~\ref{example:leadingExampleProblem}.
Fig.~\ref{figure:bwdApproxStageByStage}
describes details of one-step generation
(from the root $k=4$ to $k=3$) of the backward search
tree.
\begin{figure}[htb]
  \hspace{0.1\textwidth}
  \begin{math}
    \scriptsize
    \vcenter{
      \xymatrix@C=1.8em@R=-2em{
        {k=3}
        \ar@{<=}[r]^-*+{\text{\sf combine with fwd approx.}}
        &
        {}
        \ar@{<=}[r]^-*+{\text{\sf sense}}
        &
        {}
        \ar@{<=}[r]^-*+{\text{\sf think}}
        &
        {}
        \ar@{<=}[r]^-*+{\text{\sf act}}
        &
        {k=4}
        \\
        {\phantom{
            \begin{array}{c}
              hoge\\hoge\\hoge
            \end{array}
        }}
        \\
        *+[F]{
          \begin{array}{c}
            \cnt = 0 
            \\
              {[1 , 1.5] \cup [1.75 , 1.875]}
        \end{array}}
        &
        *+[F]{
          \begin{array}{c}
            \cnt < 1 
            \\
              {[1 , 2]}
        \end{array}}
        \ar[l]
        \\
        &&
        *+[F]{
          \begin{array}{c}
            (\Vsense \wedge \cnt < 1) \vee \neg \Vsense  
            \\
	    \\
              {[1 , 2]}    
        \end{array}}
        \ar[lu]_(.6)*++{\Vsense \mapsto \ttrue}
        \ar[ld]^(.6)*++{\Vsense \mapsto \ffalse}
        &
        *+[F]{
          \begin{array}{c}
            \Vact={\tt Acl}
            \\
	    \\
              {[1 , 2]}
        \end{array}}
        \ar[l]
        \\
        *+[F]{
          \begin{array}{c}
            \cnt = 0 \lor \cnt = 1 
            \\
            \lor \cnt = 2 \lor \cnt = 3
            \\
              {[1 , 1.2]}
        \end{array}}
        &
        *+[F]{
          \begin{array}{c}
            \True 
            \\
            \\
              {[1 , 1.2]}
        \end{array}}
        \ar[l]
        \\
        &&&&
        *+[F]{
          \begin{array}{c}
            \True
            \\
            \\
              {[1.5 , 2]}
        \end{array}}
        \ar[luu]^*+{\tt Acl}
        \ar[ldd]_*+{\tt Brk}
        \\
        *+[F]{
          \begin{array}{c}
            \cnt = 1 \lor \cnt = 2 
            \\
            \lor \cnt = 3
            \\
              {\emptyset}
        \end{array}}
        &
        *+[F]{
          \begin{array}{c}
            \cnt \geq 1 
            \\
            \\
              {[2 , 2.5]}
        \end{array}}
        \ar[l]
        \\
        &&
        *+[F]{
          \begin{array}{c}
            \Vsense \land \cnt \geq 1 
            \\
            \\
              {[2 , 2.5]}
        \end{array}}
        \ar[lu]_(.6)*++{\Vsense \mapsto \ttrue}
        \ar[ld]^(.6)*++{\Vsense \mapsto \ffalse\mathstrut}
        &
        *+[F]{
          \begin{array}{c}
            \Vact = {\tt Brk}
            \\
            \\
              {[2 , 2.5]}
        \end{array}}
        \ar[l]
        \\
        *+[F]{
          \begin{array}{c}
            \False
            \\
              {\emptyset}
        \end{array}}
        &
        *+[F]{
          \begin{array}{c}
            \False
            \\
              {[2 , 2.5]}
        \end{array}}
        \ar[l]
    }}
  \end{math}
  \caption{Generation of the backward search tree, in detail}
  \label{figure:bwdApproxStageByStage}
\end{figure}
The rightmost is the root; the four
leftmost
nodes are the direct children of the root; and the intermediate layers
are not present in  the backward search tree but are shown for
illustration.  Each of the four children corresponds to 
each possible choice of $(\sigma_{\mathsf{s}},m)$. The bottom two
children
are unsatisfiable---the intuition is that the plant's mode
at time $k=3$ cannot be $\mathtt{Brk}$ for  the
postcondition to hold. The search for a successful path will therefore
be continued from one of the two top children.


Presented in Fig.~\ref{fig:bwdSearch} is a more bird's-eye view of the
backward search tree: it shows
one possible trace of the depth-first search.
It has found a successful path
\begin{equation}\label{equation:aSuccessfulPathFoundForLeadingExample}
  \overrightarrow{(\sigma_{\mathsf{s}},m)}
  =
  \bigl\langle\,
  (x_{\mathsf{s}}\mapsto \ttrue, {\tt Acl}),\,
  (x_{\mathsf{s}}\mapsto \ffalse, {\tt Acl}),\,
  (x_{\mathsf{s}}\mapsto \ttrue, {\tt Brk}),\,
  (x_{\mathsf{s}}\mapsto \ttrue, {\tt Acl})
  \,\bigr\rangle\enspace.
\end{equation}
In the search shown in Fig.~\ref{fig:bwdSearch},  pruning has occurred at the nodes
(N1)--(N3).

\begin{figure}
  \centering
   \scalebox{.9}{  \begin{math}
    \scriptsize
    \vcenter{\xymatrix@C=1.8em@R=-.35em{
        {k=0}
        &
        {\qquad\quad k=1\qquad\quad}
        &
        {k=2}
        &
        {k=3}
        &
        {k=4}
        \\
          {\phantom{
              \begin{array}{c}
                hoge
              \end{array}
          }}
          \\
          &
          *+[F]{
            \begin{array}{c}
              \cnt = 0\\
              \emptyset
          \end{array}}
          \shifted{-3.8em}{1em}{\text{(N1)}}
          &
          *+[F]{
            \begin{array}{c}
              \False\\
                  {[0.8 , 1] \cup [1.5 , 1.75]}
          \end{array}}
          \shifted{-4.0em}{2.2em}{\text{(N3)}}
          \\
          *+[F]{
            \begin{array}{c}
              \cnt = 0 \\
                   {[0.8 , 1]}
          \end{array}}
          &
          *+[F]{
            \begin{array}{c}
              \True\\
                  {\emptyset}
          \end{array}}
          \shifted{-3.2em}{1em}{\text{(N2)}}
          &
          *+[F]{
            \begin{array}{c}
              \cnt = 0 \lor \cnt = 1 \lor \cnt = 2
              \\
                {[0.5 , 1]}    
          \end{array}}
          \ar[lu]_(.7){
            \begin{array}{c}
              \Vsense \mapsto \ttrue\\ {\tt Acl}
          \end{array}}
          \ar[l]|(.7){
            \begin{array}{c}
              \Vsense \mapsto \ffalse\\ {\tt Acl}
          \end{array}}
          \ar[ld]^(.6){
            \begin{array}{c}
              \Vsense \mapsto \ttrue\\ {\tt Brk}
            \end{array}
          }
          &
          *+[F]{
            \begin{array}{c}
              \cnt = 0
              \\
                {[1 , 1.5]}
                \\
                  {\cup\, [1.75 , 1.875]}
          \end{array}}
          \ar@/_/[lu]_*++{
            \quad
            \begin{array}{c}
              \Vsense \mapsto \ttrue\\ {\tt Acl}
          \end{array}}
          \ar[l]^*+++{
            \begin{array}{c}
              \Vsense \mapsto \ffalse\\ {\tt Acl}
          \end{array}}
          \\
          &
          *+[F]{
            \begin{array}{c}
              \cnt = 1 
              \\
                {[1 , 1.5]}
          \end{array}}
          \ar[lu]^*++{
            \begin{array}{c}
              \Vsense \mapsto \ttrue\\ {\tt Acl}
          \end{array}}
          &&&
          *+[F]{
            \begin{array}{c}
              \True 
              \\
                {[1.5 , 2]}
          \end{array}}
          \ar[lu]^*++{\begin{array}{c}
              \Vsense \mapsto \ttrue\\ {\tt Acl}
          \end{array}}
          \\
    }}
  \end{math}
}  \abovecaptionskip = 0pt
  \caption{A  bird's-eye view of the backward search tree}
  \label{fig:bwdSearch}
\end{figure}

\subsection{Synthesis of Actual Input}
\label{subsection:synthesisOfActualInput}
The second phase gives us a successful path
$\overrightarrow{(\sigma_{\mathsf{s}},m)}$; as discussed at the
beginning of~\S{}\ref{section:algorithm}, this determines the behavior 
of the whole sampled data system.
We now synthesize an actual answer
to the input synthesis problem
from the path
$\overrightarrow{(\sigma_{\mathsf{s}},m)}$. 
Theoretically it is possible (Prop.~\ref{proposition:correctnessOfBwdSearchTree});
it is moreover computationally cheap,
 using a CAS like Mathematica.

We describe the procedure by example. For
Example~\ref{example:leadingExampleProblem},  the second phase
gives a successful path
in~(\ref{equation:aSuccessfulPathFoundForLeadingExample}), from which 
we obtain a refinement 
 of the pre-CP-condition
\begin{math}
  \BS
  \bigl(
  \overrightarrow{(\sigma_{\mathsf{s}},m)}
  \bigr)
  =
  (\cnt = 0,\, [0.8, 1]) 
\end{math}
(the leftmost node
in Fig.~\ref{fig:bwdSearch}).

\vspace*{-.8em}
\begin{itemize}
\item 
  \textbf{(Choosing an initial state)}
  By Prop.~\ref{proposition:correctnessOfBwdSearchTree}
   any $(\sigma_0, v_0)$ such that
  $(\sigma_0, v_0)\models  \BS
  \bigl(
  \overrightarrow{(\sigma_{\mathsf{s}},m)}
  \bigr)$ 
  admits a desired input sequence. 
  Let us say
 $\sigma_{0}(\cnt) = 0$ and $v_{0}
  := 0.9$.
\item 
  \textbf{(Running the plant)}
  It is crucial that the behavior of the plant is completely determined
  now, given the initial P-state $v_{0}$ and the sequence of
  modes $\langle m^{(0)},\dotsc, m^{(T-1)}\rangle$  extracted from the
   path $\overrightarrow{(\sigma_{\mathsf{s}},m)}$. In the
  current example the plant dynamics is as follows:
  \begin{math}
    0.9 \xrightarrow[{\tt Acl}]{} 1.45 \xrightarrow[{\tt Brk}]{} 0.95 \xrightarrow[{\tt Acl}]{} 1.475 \xrightarrow[{\tt Acl}]{} 1.7375
  \end{math}.
\item 
  \textbf{(Synthesis of input)}
  For each moment $k$, we now know the plant state $v^{(k)}$ 
  and the sensor output $\sigma_{\mathsf{s}}^{(k)}\in \B^\VsenseSet$;
      the latter is
  extracted from the
   path $\overrightarrow{(\sigma_{\mathsf{s}},m)}$. We choose
  input $i_{k}$ so that it, combined with $v^{(k)}$, gives
  the sensor output as specified by
  $\sigma_{\mathsf{s}}^{(k)}$. 

  For example let us pick $i_{2}$. Now
  $\sigma^{(2)}(\Vsense)=(v+i\ge 1?)=\ffalse$ and $v^{(2)}=0.95$; we choose $i_{2}\in
  I=[-0.2, 0.2]$ so
  that
  $v^{(2)}+i_{2}=0.95+i_{1}<1$; say $i_{2}=0$. In 
  implementation we let the 
  $\mathtt{FindInstance}$
  function of Mathematica do this job.
\end{itemize}
Overall, we obtain the following run from the
pre-CP-condition
to the post-CP-condition. This gives an answer
\begin{math}
    (\cnt\mapsto 0, v=0.9) \xrightarrow{0.1} (1, 1.45)
  \xrightarrow{0.0} (
  2, 0.95) \xrightarrow{0.0}
  (
  0, 1.475) \xrightarrow{0.0} (
  1, 1.7375)
\end{math}
 to the input synthesis
problem in Example~\ref{example:leadingExampleProblem}.

\section{Implementation, Optimization and Experiments}\label{section:optimizationAndImpl}


Our prototype implementation has a front-end written in OCaml which in
particular  implements inferences in program logic
(Def.~\ref{definition:weakestPreconditionAndStrongestPostcondition}). Mathematica
is used for simplifying arithmetic formulas and inequalities,
as well as for picking a value under a certain assumption.
We also use
MATLAB for numerically solving ODEs.



Our implementation is currently restricted to one-dimensional plants
($n=1$). From time to time we have to calculate the evolution of an interval 
according an ODE (like $\Execplant(p_m, X)$
in~(\ref{equation:defOfStagesOfOneFA}) for a set $X$); such calculation is done
by the method of~\cite{DBLP:conf/sefm/EggersRNF11}.


\paragraph{Optimization Techniques}
We further employ the following techniques for speedup.
We note that the none of these affect
correctness
(Prop.~\ref{proposition:correctnessOfBwdSearchTree}) of our algorithm.
\begin{itemize}
 \item 
 \textbf{(Truncation of forward approximation)}
 In the forward approximation phase, 
 a problem is that an approximant $\kFA(\Phi_{\init}, X_{\init})$
 can grow exponentially as $k$ grows---as hinted  already
 in~(\ref{equation:fwdApproxPicture}). Such explosion of
 approximants slows down not only the forward approximation phase,
 but also the backward search phase.
 Moreover such a
 big approximant tends  not to contribute a lot to
 pruning branches.

 To avert this we truncate forward approximation under 
 certain circumstances. Specifically we stop calculating C-conditions
 when the approximated C-condition has become compatible with any 
 choice of modes---a sign of the C-condition  no longer
 contributing to pruning. Currently 
 such truncation is implemented only for C-conditions; but it should
 also be possible for P-conditions, e.g.\ by merging intervals in~(\ref{equation:fwdApproxPicture}).

 \item 
 \textbf{(Prioritization in search)}
 In the backward (depth-first) search phase, 
 we can have multiple children from which to pick.
 Besides randomized picks, we have the following prioritization 
 strategies.
 In the \emph{by-volume prioritization}, we estimate the volume of the
 P-condition (i.e.\ a region in $\R^{n}$) of each child, and pick one
 with the biggest. In the \emph{by-robustness prioritization}, in
 contrast, we pick the child whose P-condition is the closest to the
 ``center'' of the forward-approximated P-condition. In other words, the
 picked child is the one with a P-condition that intersects
 with the forward-approximated P-condition \emph{in the most robust manner}. 
 This robustness-driven optimization is much like in S-Taliro~\cite{DBLP:conf/tacas/AnnpureddyLFS11}.



\end{itemize}





\paragraph{Experiments}
We used 
Mathematica 9.0.1
and MATLAB 8.1.0 (for Linux x86, 64-bit),
on ThinkPad T530
with Intel Core i7-3520M 2.90GHz CPU
with 3.7GB memory.

The first table below
 shows the result of our prototype
implementation applied to the problem in
Example~\ref{example:leadingExampleProblem}, with a varying number $T$
of steps. All the times are in seconds. 
The rows correspond to different prioritization
strategies,
and whether truncation of forward approximation is enabled. For random
prioritization the experiment was repeated 50 times and the average is 
shown,  together with the standard deviation.
%
%
From the results we can see that forward approximation truncation is
very
effective as the problem becomes larger, on the one hand. 
On the other hand, no clear comparative
advantage of any of the three prioritization strategies is observed.


The second table below
 presents the breakdown of
two cases (both $T=100$) from  the first table, into the three phases of
the algorithm, together with the number of backtracks in a search. While
truncation causes more backtracks (this is because less information is
passed to the backward search phase), we see that the speed of both of
the first two phases are greatly improved thanks to simpler approximants.
%
%

We also applied our  implementation to the original problem
in
Example~\ref{example:TYTexample}. 
It  successfully solved the problem in
638.968 seconds.

Overall, our experiments so far are limited to  examples of a specific
structure: namely, a counter in the controller,  incremented or reset
to $0$ every second, causes the change of  modes of the plant. This
structure however is a commonly used one in industry (see Example~\ref{example:CountAndBrake}); and its discrete
nature (the counter takes an integer value that can be fairly large)
becomes a challenge in many approaches to verification, testing or input
synthesis.  
The experimental results seem to  suggest that
our program logic based approach is promising 
in coping with this kind of challenges.



\begin{center}
\scalebox{.9}{\small
   \begin{tabular}{l|c||r|r|r|r|r}
    prioritization &  truncation       &  $T=10$ &   $T=20$ &   $T=30$ &   $T=100$ &    $T=1000$\\  \hline\hline
    random &           & 3.228 &  9.849 & 18.464 & 108.437 $\pm$ 37.574 & No answer\\  \hline
    by volume&        & 3.633 & 10.197 & 14.332 & 115.311              & No answer\\  \hline
    by robustness &        & 3.072 & 11.082 & 22.231 &  68.464              & No answer\\  \hline
    random & on  & 3.409 & 11.314 & 20.068 &  54.132 $\pm$ 31.953              &   377.361 $\pm$ 80.392\\  \hline
    by volume & on & 3.689 &  9.289 & 12.323 &  38.425              &   445.784\\  \hline
    by robustness & on & 3.552 & 20.702 & 41.443 &  38.803              &   245.661\\  
  \end{tabular}
}

  \vspace{.5em}
\scalebox{.9}{\small
  \begin{tabular}{l|c||r|r|r|r|r} 
    prioritization & truncation  & fwd.\ approx. &  bwd.\ search & synthesis & total & num.\ of backtracks\\  \hline\hline
    by volume &  & 36.622 & 76.568 & 2.121 & 115.311 & 140\\  \hline
    by volume & on & 15.118 & 21.565 & 1.743 &  38.425 & 176
  \end{tabular}}
\end{center}

\vspace{-2em}

\bibliographystyle{eptcs}
\bibliography{relatedWork}

\newpage

\appendix

\section{Auxiliary Definitions and  Lemmas}

\begin{mydefinition}[semantics $\Denote{\place}$ of $\ImpCtrl$]
  \label{definition:semanticsOfImpCtrl}
  Let $\State$ be the set of \emph{valuations}, that is, 
  \begin{equation}\label{equation:valuation}
    \State = \bigl\{ \sigma:\Var\to\R \cup \B \cup \Modes \;\bigl|\bigr.\; \sigma(\VthinkSet) \subseteq \R, \sigma(\VsenseSet) \subseteq \B, \sigma({\Vact}) \in \Modes \bigr\}\enspace.
  \end{equation} 
  
  For each expression $e$ of $\ImpCtrl$, their \emph{semantics} $\Denote{e}$
  is defined in the following standard way.
  For $a \in \AExp$, $\Denote{a} : \State \to \R$ is defined by
  \begin{displaymath}
    \begin{array}{rllrllrll}
      \Denote{r}(\sigma)& = & r\enspace,\quad
      &
      \Denote{\Vthink}(\sigma)& = & \sigma(\Vthink) \enspace,\quad
      &
      \Denote{a_1 \aop a_2}(\sigma)& = & \Denote{a_1}(\sigma) \aop
      \Denote{a_2}(\sigma)\enspace.
    \end{array} 
  \end{displaymath}
  For $b \in \BExp$, $\Denote{b} : \State \to \{\ttrue, \ffalse\}$ is
  defined by
  \begin{displaymath}
    \begin{array}{rlll}
      \Denote{b_1 \vee b_2}(\sigma)& = & \Denote{b_1}(\sigma) \vee
      \Denote{b_2}(\sigma)\quad\text{and similarly for $\neg, \wedge, \True$
        and $\False$;}
      \\
      \Denote{\Vsense}(\sigma)& = & \sigma(\Vsense)\enspace; \quad\text{and}\qquad
      \Denote{a_1 \rop a_2}(\sigma) =  \Denote{a_1}(\sigma) \rop \Denote{a_2}(\sigma)\enspace.
    \end{array} 
  \end{displaymath}
  For $c \in \Cmd$, $\Denote{c} : \State \to \State$ is defined by
  \begin{displaymath}
    \begin{array}{c}
      \Denote{\Skip}(\sigma) =  \sigma\enspace,
      \quad
      \Denote{\Vthink := a}(\sigma) =  \sigma[\Vthink \mapsto
        \Denote{a}(\sigma)]   \enspace,
      \quad
      \Denote{\Vact := m_i}(\sigma) =  \sigma[\Vact \mapsto m_i]\enspace,\\
      \Denote{c_1;c_2}(\sigma) =  \Denote{c_2} \bigl(
      \Denote{c_1}(\sigma)\bigr)\enspace, \quad
      \Denote{\If{b}{c_1}{c_2}}(\sigma)  =  
      \begin{cases}
        \Denote{c_1}(\sigma) & \mathrm{if}\; \Denote{b}(\sigma) = \ttrue\\
        \Denote{c_2}(\sigma) & \mathrm{if}\; \Denote{b}(\sigma) = \ffalse.
      \end{cases}
    \end{array} 
  \end{displaymath}  
  Here $f[x_0 \mapsto y_0]$ denotes \emph{function update}: the function
  $f[x_0 \mapsto y_0]$ carries $x_{0}$ to $y_{0}$ and acts as $f$ on the
  other input.
\end{mydefinition}

\begin{mydefinition}[semantics of $\AssnCtrl$]
  \label{definition:semanticsOfAssnCtrl}
  We define the semantics of $a \in \AExp$ of $\AssnCtrl$ as a function $\Denote{a}:\State \times \R^{\Var'} \to \R$.
  \begin{displaymath}
    \begin{array}{rlllrlll}
      \Denote{r}(\sigma, \gamma)& = & r \enspace ,& \quad &
      \Denote{\Vthink}(\sigma, \gamma)& = & \sigma(\Vthink) \enspace , & \\
      \Denote{v'}(\sigma, \gamma)& = & \gamma(v') \enspace, & \quad &
      \Denote{a_1 \aop a_2}(\sigma)& = & \Denote{a_1}(\sigma, \gamma) \aop \Denote{a_2}(\sigma, \gamma)&\\
    \end{array} 
  \end{displaymath}
  For $m \in \MExp$, its semantics is a function $\Denote{m}:\State
  \times \R^{\Var'} \to \Modes$ defined by $\Denote{m_i}(\sigma, \gamma) =
  m_i$ and $\Denote{\Vact}(\sigma, \gamma) = \sigma(\Vact)$. 
  Finally for formulas, the semantics of $\Phi \in \Fml$ is given by the
  relation $\models$ between $\State \times \R^{\Var'}$ and $\Fml$
  defined as follows. $\Assn{\sigma, \gamma}{\True}$, 
  $\NotAssn{\sigma, \gamma}{\False}$,  and
  \begin{displaymath}
    \begin{array}{rl}
      \Assn{\sigma, \gamma}{\Vsense} \;\Defiff\;& \sigma(\Vsense) = \ttrue \enspace ,\\
      \Assn{\sigma, \gamma}{a \rop a'} \;\Defiff\; & \Denote{a}(\sigma, \gamma) \rop \Denote{a'}(\sigma, \gamma) \enspace, \quad\\
      \Assn{\sigma, \gamma}{m = m'} \;\Defiff\; & \Denote{m}(\sigma, \gamma) = \Denote{m'}(\sigma, \gamma) \enspace , \quad\\
      \Assn{\sigma, \gamma}{\Phi_1 \vee \Phi_2} \;\Defiff\; &\Assn{\sigma, \gamma}{\Phi_1} \text{ or } \Assn{\sigma, \gamma}{\Phi_2} \quad
      \text{(similarly for $\neg$ and $\wedge$)},\\
      \Assn{\sigma, \gamma}{\forall v' \in \R.\Phi} \;\Defiff\; &\Assn{\sigma, \gamma[v' \mapsto r]}{\Phi} \text{ for any } r \in \R \quad
      \text{(similarly for $\exists$)}.\\
    \end{array} 
  \end{displaymath}
  We write $\Assn{\sigma}{\Phi}$ if $\Assn{\sigma, \gamma}{\Phi}$ for every $\gamma \in \R^{\Var'}$.
\end{mydefinition}

The next observation follows immediately from
Def.~\ref{definition:STAHS} and~\ref{definition:conditionOnSTAHS}.
\begin{mylemma}\label{lemma:unsatisfiableCondition}
  $(\Phi, X)$ is unsatisfiable if and only if $\Phi$ is an
  unsatisfiable formula (i.e.\ logically equivalent to $\False$) or $X =
  \emptyset$. \myqed
\end{mylemma}

\section{Omitted Proofs}
\subsection{Proof of
  Prop.~\ref{proposition:1stepForwardApproxCorrectness}}
\begin{mylemma}\label{lemma:componet1stepForwardApproxCorrectness}
  Let  $i \in I$, $\sigma \in \State$ and $x \in \R^n$. Assume
  $(\sigma,x)\models (\Phi,X)$. Then: 
  \begin{equation}
    \begin{array}[d]{rlrlrl}
      \Sense(\sigma, x, i) &\;\models\;
      \PreOneFA_{\Sense}(s(i,x))(\Phi,X)\enspace,
      &\quad
      \Think(\sigma, x)    &\;\models\;
      \PreOneFA_{\Think}(\Phi,X)\enspace,
      \\
      \Act(\sigma, x)      &\;\models\; \PreOneFA_{\Act}(\sigma(x_{\mathsf{a}}))(\Phi,X)\enspace.\\
    \end{array}
  \end{equation}
\end{mylemma}
\begin{myproof}
  For $\Sense$, 
  it is easily shown that $\sigma[\Vsense \mapsto s(x,i)] \,\models\, \SPsyntax{x_{\mathsf{s}} := \sigma_{\mathsf{s}}(x_{\mathsf{s}})}{\Phi}$ by induction on $\Phi$.
  Moreover $x \in X \cap s^{-1}(s(x,i))$ by Notation~\ref{notation:sInverse}.
  Therefore we have $\Sense(\sigma, x, i) \,\models\, \PreOneFA_{\Sense}(s(i,x))(\Phi,X)$.
  
  For $\Think$, the claim is obvious from
  Prop.~\ref{proposition:IndeedWeakestPreconditionAndStrongestPostcondition}.

  For $\Act$, we trivially have $\sigma \,\models\, x_{\mathsf{a}} = \sigma(x_{\mathsf{a}})$. Therefore
  \begin{displaymath}
    (\sigma, \Execplant(p_{\sigma(x_{\mathsf{a}})}, x)) \,\models\, (\Phi \wedge x_{\mathsf{a}} = \sigma(x_{\mathsf{a}}), \Execplant(p_{\sigma(x_{\mathsf{a}})},X)) 
  \end{displaymath}
  follows immediately from the assumption.
  \myqed
\end{myproof}

\begin{myproof} (Of Prop.~\ref{proposition:1stepForwardApproxCorrectness})
  It is sufficient to show for $k=1$; the general case follows by induction.
  Let
  $i \in I$,
  $(\sigma,x) \xrightarrow{i} (\sigma',x')$ be a run of $\Sys$, 
  and $(\sigma,x)\models (\Phi,X)$. We need to show
  \begin{displaymath}
    (\sigma',x')
    \,\models\,
    \OneFA(\Phi,X)\enspace.
  \end{displaymath}
  That is obvious because we  obtain the following from Lemma.~\ref{lemma:componet1stepForwardApproxCorrectness}.
  \begin{equation}
    (\sigma',x') \,\models\, \PreOneFA(s(i,x),
    \sigma'(x_{\mathsf{a}}))(\Phi,X) \enspace.
    \tag*{\myqed}
  \end{equation}
  \auxproof{  Note here that $x$ is the \emph{pre} P-state and $\sigma'$ is the \emph{post} C-state,
    which are correct because of the sense-think-act cycle 
    (\ref{equation:structureOfSTAHS}).}
\end{myproof}


\subsection{Proof of Prop.~\ref{proposition:correctnessOfBwdSearchTree}}
\begin{mylemma}\label{lemma:componentCorrectnessOfBwdSearchTree}
  The following three properties hold.
  \begin{displaymath}
    \begin{array}{rl}
      (\sigma, x) \,\models\, \PreBS_{\Sense}(\sigma_{\mathsf{s}})(\Phi, X)
      &\implies \Sense(\sigma, x, i) \,\models\, (\Phi, X) \;\text{for some}\;i \in I
      \\
      (\sigma, x) \,\models\, \PreBS_{\Think}(\Phi, X) &\implies \Think(\sigma, x) \,\models\, (\Phi, X)
      \\
      (\sigma, x) \,\models\, \PreBS_{\Act}(m)(\Phi, X) &\implies \Act(\sigma, x) \,\models\, (\Phi, X)
    \end{array}
  \end{displaymath}
  It follows that: if
  \begin{math}
    (\sigma,x) \,\models\, \PreBS(\sigma_{\mathsf{s}}, m)(\Phi, X)
  \end{math},
  then there exist input $i\in I$ and a CP-state $(\sigma',x')$ for which we have
  $(\sigma, x) \xrightarrow{i} (\sigma', x') $ and
  $(\sigma', x') \,\models\, (\Phi, X)$.
\end{mylemma}
\begin{myproof}
  For $\Sense$, 
  it follows from the assumption that
  $\sigma \,\models\, \Phi[\sigma_{\mathsf{s}}(\Vsense)/\Vsense]$ and $x \in X \cap s^{-1}(\sigma_{\mathsf{s}})$.
  Then we have $(\sigma[\Vsense \mapsto \sigma_{\mathsf{s}}(\Vsense)], x) \,\models\, (\Phi, X)$.
  Moreover, since $x\in s^{-1}(\sigma_{\mathsf{s}})$ there exists some $i \in I$  such that $s(x,i) = \sigma_{\mathsf{s}}$.
  For this choice of $i$ we have $\Sense(\sigma, x, i) = (\sigma[\Vsense \mapsto \sigma_{\mathsf{s}}(\Vsense)], x)$.

  For $\Think$, we have $\sigma \,\models\, \WPsyntax{c}{\Phi}$ and $x \in X$ from the assumption.
  Therefore $\bigl(\Denote{c}(\sigma), x\bigr) \,\models\, (\Phi, X)$
  from Prop.~\ref{proposition:IndeedWeakestPreconditionAndStrongestPostcondition}.

  For $\Act$,  we have
  $\sigma \,\models\, \Phi \wedge x_{\mathsf{a}} = m$ and $x \in \Execplant(\rev(p_m), X)$ from the assumption.
  Then we have  $\sigma(x_{\mathsf{a}}) = m$, 
  hence 
  $x \in \Execplant(\rev(p_{\sigma(x_{\mathsf{a}})}), X)$, that is,
  $x$ is reached from the region $X$ by running
  $p_{\sigma(x_{\mathsf{a}})}$
  with time reversed. From this the claim
  $(\sigma, \Execplant(p_{\sigma(x_{\mathsf{a}})}, x)) \,\models\, (\Phi,
  X)$
  follows.
  \myqed

\end{myproof}

\begin{myproof} (Of Prop.~\ref{proposition:correctnessOfBwdSearchTree})
  For soundness, 
  first we observe that $(\sigma_0, x_0) \,\models\, (\Phi_{\init},
  X_{\init})$. This is because
  $  \BS
  \bigl(
  \overrightarrow{(\sigma_{\mathsf{s}},m)}
  \bigr)
  $ implies 
  $
  (\Phi_{\init}, X_{\init})=0\text{-}\FA(\Phi_{\init}, X_{\init})
  $ by Def.~\ref{definition:backwardSearchTree}; see in particular~(\ref{equation:childInBackwardSearchTree}).

  Starting from $(\sigma_0, x_0)$,
  we can repeatedly apply Lem.~\ref{lemma:componentCorrectnessOfBwdSearchTree}
  to obtain 
  input $i_{0},\dotsc, i_{T-1}$ 
  and CP-states $(\sigma_1, x_1),\dotsc,(\sigma_T, x_T)$ such that:
  $(\sigma_0, x_0) \xrightarrow{i_0} \cdots \xrightarrow{i_{T-1}}
  (\sigma_T,x_T)$;
  and 
  $(\sigma_{k},x_{k})\models \BS\bigl(
  (\sigma_{\mathsf{s}}^{(T-1)},m^{(T-1)}),
  \dotsc,
  (\sigma_{\mathsf{s}}^{(k)},m^{(k)})
  \bigr)$ for each $k\in[0,T]$.
  Then in particular 
  $(\sigma_{T},x_{T})\models \BS(\varepsilon)=(\Phi_{\final}, X_{\final})
  $ where $\varepsilon$ denotes the empty sequence. This means that
  $(\sigma_{0},x_{0})$ and  $i_{0},\dotsc, i_{T-1}$ qualify as an answer.


  For completeness, 
  let $(\sigma_0,x_0) \xrightarrow{i_0} \cdots \xrightarrow{i_{T-1}} (\sigma_T,x_T)$ be a run of $\Sys$.
  We can take 
  \begin{equation}
    \overrightarrow{(\sigma_{\mathsf{s}},m)}:=
    \bigl\langle\,
    \bigl(s(x_{T-1}, i_{T-1}), \sigma_T(x_{\mathsf{a}})\bigr),\,
    \dotsc,\,
    \bigl(s(x_{0}, i_{0}),\sigma_1(x_{\mathsf{a}})\bigr)
    \,\bigr\rangle
    \enspace.
    \tag*{\myqed}
  \end{equation}

\end{myproof}

\subsection{Proof of Lem.~\ref{lemma:pruningIsCorrect}}
\begin{myproof}
  Assume the node $(\Phi, X)$ at the position
  $\overrightarrow{(\sigma_{\mathsf{s}},m)}$ has an unsatisfiable label.
  Consider its child at the position
  $\overrightarrow{(\sigma_{\mathsf{s}},m)}(\sigma_{\mathsf{s}}', m')$,
  for arbitrary $\sigma_{\mathsf{s}}' \in \B^\VsenseSet$ and $m' \in \Modes$.

  We know $\Phi = \False$ or $X = \emptyset$ from Lem.~\ref{lemma:unsatisfiableCondition}.
  In case $\Phi = \False$, we easily see that
  \begin{displaymath}
    \CtrlCond( \PreBS(\sigma_{\mathsf{s}}, m)(\Phi,X)) = \False \enspace.
  \end{displaymath}
  In case $X = \emptyset$, similarly
  \begin{equation}
    \PlantCond( \PreBS(\sigma_{\mathsf{s}}, m)(\Phi,X)) = \emptyset \enspace.
    \tag*{\myqed}
  \end{equation}
\end{myproof}

\end{document}